\documentclass[preprint,3p]{elsarticle}

\usepackage{lineno,hyperref}
\modulolinenumbers[5]

\journal{JSV}

\usepackage[usenames,dvipsnames]{xcolor}

\usepackage{graphicx}        

\usepackage{natbib}

\usepackage{amsmath}
\usepackage{amsfonts}
\usepackage{amssymb}

\usepackage{subfigure}
\usepackage{xspace}
\newcommand{\bs}{\boldsymbol}

\newcommand{\etal}{et al.\xspace}
\newcommand{\ie}{i.\,e.\xspace}
\newcommand{\eg}{e.\,g.\xspace}
\newcommand{\cf}{cf.\xspace}
\newcommand{\fp}{\,.}
\newcommand{\fk}{\,,}

\newcommand{\mrm}[1]{\mathrm{#1}} 

\newcommand{\ee}{\mrm{e}}
\newcommand{\ii}{\mrm{i}}

\newcommand{\nc}{\newcommand}
\newcommand{\rnc}{\renewcommand}
\nc{\tra}{^{\mathrm T}}
\nc{\VINES}{vibro-impact NES\xspace}
\nc{\sref}[1]{Sect.~\ref{sec:#1}}
\nc{\srefs}[1]{Sect.~\ref{sec:#1}}
\nc{\srefo}[1]{\ref{sec:#1}}
\nc{\ssref}[1]{Subsect.~\ref{sec:#1}}
\nc{\ssrefs}[1]{Subsect.~\ref{sec:#1}}
\nc{\ssrefo}[1]{\ref{sec:#1}}
\nc{\eref}[1]{Eq.~\ref{eq:#1}}
\nc{\erefs}[1]{Eqs.~\ref{eq:#1}}
\nc{\erefo}[1]{\ref{eq:#1}}
\nc{\fref}[1]{Fig.~\ref{fig:#1}}
\nc{\frefs}[1]{Figs.~\ref{fig:#1}}
\nc{\frefo}[1]{\ref{fig:#1}}
\nc{\tref}[1]{Tab.~\ref{tab:#1}}
\nc{\fig}[4][tbh]{
\begin{figure}[#1]
\centering
\includegraphics[width=#4\textwidth]{figs/#2}
\caption{#3\label{fig:#2}}
\end{figure}}
\nc{\ea}[1]{
\begin{eqnarray}
#1\end{eqnarray}}
\rnc{\matrix}[2]{\left[\!\!\begin{array}{#1}
	#2\end{array}\!\!\right]}
\rnc{\vector}[1]{\matrix{c}{#1}}
\nc{\dd}{\mathrm{d}}
\usepackage{color}
\nc{\COMMENT}[1]{\textcolor{red}{#1}}
\nc{\COMMENTth}[1]{\textcolor{blue}{#1}}
\usepackage{wasysym}
\usepackage{tikz} 
\nc{\circled}[1]{(#1)}
\nc{\pmpar}{{}_{\scriptscriptstyle (}\!\pm\!{}_{\scriptscriptstyle )}} 

\nc{\qahat}{\hat q_{\mathrm a}} 
\nc{\q}{q} 
\nc{\amp}{a} 
\nc{\ares}{a_{\mathrm{res}}}
\nc{\aresnoabs}{a_{\mathrm{res}}^{\mathrm{no abs}}}
\nc{\psiA}{\Delta} 
\nc{\psiE}{\theta} 
\nc{\phic}{\varphi_\mrm{c}} %
\nc{\psiEx}{\varphi_\mrm{ex}} %
\nc{\pmr}{\varepsilon_\mrm{phy}} 
\nc{\mmr}{\varepsilon_\mrm{mod}} 
\nc{\mm}{\bs}
\nc{\ma}{m_\mrm a} 
\nc{\cor}{r} 
\nc{\gap}{g} 
\nc{\D}{D} 
\nc{\fex}{{\hat f}_\mrm{ex}} 
\nc{\vc}{v_\mrm{c}} 
\nc{\tc}{T_\mrm{c}} 
\nc{\fc}{\hat{f}_\mrm{c}} 
\nc{\compr}{\alpha} 
\nc{\tcO}{T_\mrm{c,0}} 
\nc{\fcO}{\hat{f}_\mrm{c,0}} 
\nc{\comprO}{\alpha_0} 
\nc{\sd}{\sin\Delta} 
\nc{\cd}{\cos\Delta} 

\usepackage{stfloats}
\fnbelowfloat 
\usepackage[noprefix]{nomencl}
\makenomenclature
\usepackage{cancel}
\usepackage{cases}

\makeindex             

\begin{document}

\begin{frontmatter}
\title{Predictive design of impact absorbers for mitigating resonances of flexible structures using a semi-analytical approach}
\author[addressILA]{Timo Theurich}
\author[addressDMSE]{Alexander F. Vakakis}
\author[addressILA]{Malte Krack}

\address[addressILA]{Institute of Aircraft Propulsion Systems, University of Stuttgart, Pfaffenwaldring 6, 70569 Stuttgart, Germany\\ theurich@ila.uni-stuttgart.de, krack@ila.uni-stuttgart.de}
\address[addressDMSE]{Department of Mechanical Science and Engineering, University of Illinois at Urbana – Champaign, 1206 W. Green Str., Urbana, IL 61801, USA\\ avakakis@illinois.edu}

\begin{abstract}
Analytical conditions are available for the optimum design of impact absorbers (or Vibro-Impact Nonlinear Energy Sinks) for the special case where the host structure is well described as a rigid body.
Accordingly, the analysis relies on the assumption that the impacts cause immediate dissipation in the contact region, which is modeled in terms of a coefficient of restitution (Newton's impact law) that is assumed to be known and fixed.
When a flexible host structure is considered instead, the impact absorber not only dissipates energy at the time instances of impact, but, and perhaps equally important, it inflicts nonlinear energy scattering between structural modes. Hence, it is crucial to account for such nonlinear energy transfers yielding energy redistribution within the modal space of the structure.
In the present work, we develop a design approach for flexible host structures.
We consider the case of a forced excitation near primary resonance with a well-separated mode.
We demonstrate that the time scales of the impact and the resonant vibration can be decoupled.
On the long time scale, the dynamics of the host structure can be properly reduced to the fundamental harmonic of the resonant mode.
A light impact absorber responds to this enforced motion, and we show that the conventional Slow Invariant Manifold of the dynamics is recovered for the commonly considered regime of two symmetric impacts per period.
On the short time scale of the impact dynamics, the contact mechanics and elasto-dynamics must be finely resolved.
We show that it is sufficient to run a numerical simulation of a single impact event with representative pre-impact velocity.
From this short-time simulation, we derive an effective modal coefficient of restitution and the properties of the contact force pulse, needed to approximate the behavior on the long time scale.
We derive a closed-form expression of the periodic resonant response and establish that the design problem can be reduced to four dimensionless parameters.
We demonstrate the approach for the numerical example of a cantilevered beam with a spherical impact absorber.
We conclude that the proposed semi-analytical procedure enables deep qualitative understanding of the problem and, at the same time, yields a quantitatively accurate prediction of the optimum design.
\end{abstract}


\end{frontmatter}

\section{Introduction} \label{sec:intro}
An impact absorber is a light nonlinear attachment designed to undergo impacts with a host structure whose vibrations are to be mitigated.
For instance, the absorber can have spherical form and can be placed inside a cavity of the host structure.
Other common names for the same device are impact damper, impulse mistuner \cite{Hartung2016} and Vibro-Impact Nonlinear Energy Sink (VI-NES) \cite{vaka2008b,Lamarque2011,Gendelman.2012}.
Apparently, the device dates back to the 1940s \cite{Lieber.1945} and has been frequently used since then.
A characteristic feature of Nonlinear Energy Sinks is that they can engage in resonance with arbitrary modes of the structure to which they are attached over broad frequency ranges. This is an important advantage over passive tuned vibration absorbers / tuned mass dampers which are narrowband devices and their effect is sensitive to structural changes \cite{vaka2008b}.
The impact absorber has been shown to effectively mitigate free (unforced) vibrations under impulsive (shock, seismic) excitation \cite{vaka2008b,AlShudeifat.2013}, resonantly-driven vibrations \cite{Bapat.1985,Gendelman.2015,Theurich.2019} and self-excited vibrations \cite{Chatterjee.1996,Asfar.2005,mueller.2018}.
\\
An important design parameter of impact absorbers is the clearance between absorber and host structure.
In \fref{schematicEfficacyCurve}, the resonant response is schematically illustrated as function of the clearance.
Throughout this work, we define as resonant response the steady-state response with the maximum displacement reached within the excitation frequency range near the considered resonance.
If the clearance is large, no or seldom impacts occur, and the system responds as if the absorber was not present.
In the other limit, if the clearance becomes negligible compared to the vibration amplitude, the absorber is in almost permanent contact, and the system responds as if the absorber was fixed to the structure (rigidly attached mass).
In between these extreme cases, the resonant response can be considerably reduced compared to the case without absorber.
The reduction effect is more pronounced for larger absorber masses.
Using multiple absorbers with specific clearances is a possibility to improve the robustness with respect to the excitation level \cite{Li.2017,Boroson.2017,Vaurigaud.2011,Fang.2020,Qiu.2019}.
Interestingly, the character of the resonant response switches at the optimum:
For slightly smaller clearances, the resonant response is periodic with two symmetrical impacts per period.
For slightly larger clearances, the resonant response is strongly modulated and the pattern of the impacts changes and can become even irregular.
It is useful to characterize the resonant response in terms of maximum and mean amplitude (\fref{schematicEfficacyCurve}).
Note that if the host structure is not a single-degree-of-freedom system, a small chaotic perturbation remains for sufficiently light damping in the nominally periodic regime as demonstrated numerically in \cite{Theurich.2019} and experimentally in \cite{Li.2016b}.
\begin{figure}[t!]
	\centering
	\includegraphics[width=0.6\textwidth]{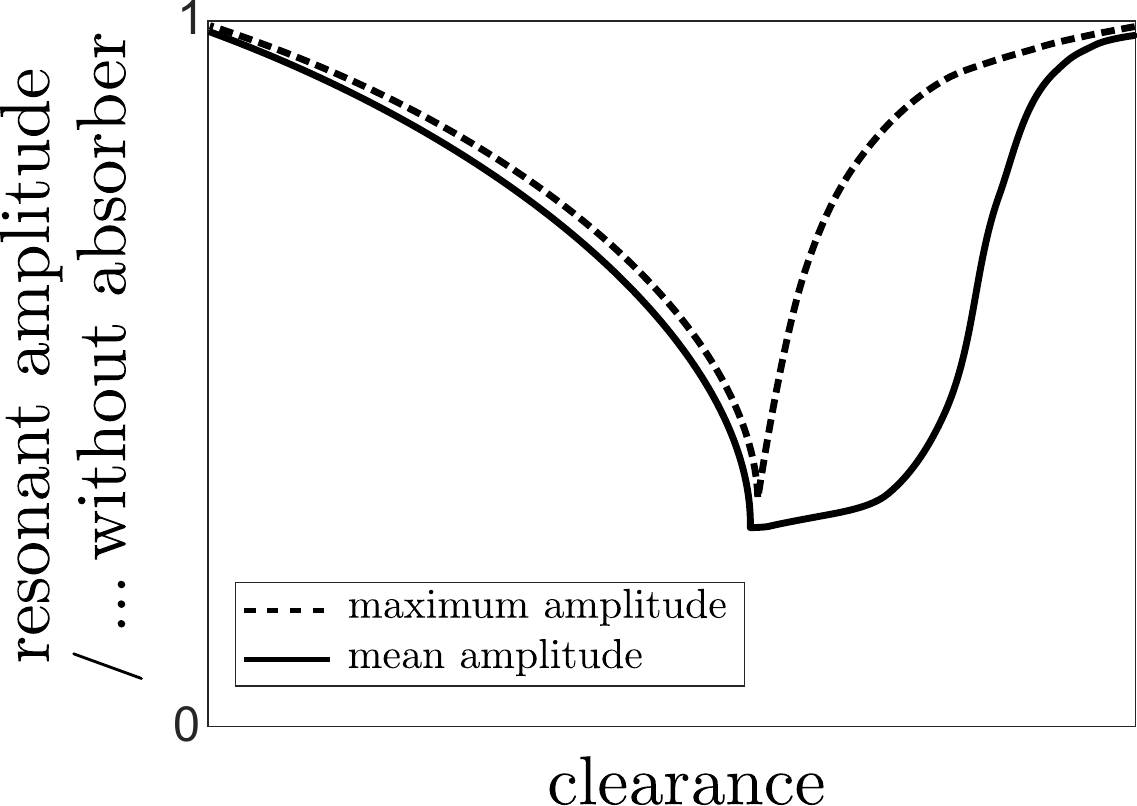}
	\caption{Schematic illustration of the resonant response (normalized by the response without absorber) as function of the clearance.}
	\label{fig:schematicEfficacyCurve}
\end{figure}
\\
The approach developed by Gendelman \cite{Gendelman.2012} permits the analytical study of a spring-mass-oscillator with an impact absorber if the impacts are modeled with a known and fixed coefficient of restitution (Newton's impact law).
Using the method of multiple scales, it is shown that the system's center of mass oscillates harmonically to first-order approximation, provided that the absorber and the linear viscous damping are light.
Assuming that the absorber synchronizes with a 1:1 resonance, with two symmetric impacts per period, the absorber undergoes a sawtooth oscillation.
In this regime, a so-called Slow Invariant Manifold (SIM) can be attributed to the (slow) dynamics, relating the amplitudes of the primary oscillator and the absorber, the clearance and the coefficient of restitution.
It can be shown that the SIM has two branches, and only a certain part of one branch is stable.
Further, it can be established that below a certain amplitude, no stable periodic oscillation exists.
Under weak harmonic excitation near primary resonance, one can derive an algebraic equation which relates the response amplitude, damping, excitation level and excitation frequency.
This equation is given explicitly in \cite{Gendelman.2015} for the case of forced excitation, and in \cite{Qiu.2019} for the case of base excitation.
The series of papers by Seguy, Berlioz, Michon \etal \cite{Gourc.2015,Li.2016,Li.2016b,Li.2017,Li.2017b,Qiu.2019} relies on the above described analytical approach developed by Gendelman \cite{Gendelman.2012}.
An experimental confirmation of periodic and Strongly Modulated Response (SMR) regimes is given in \cite{Gourc.2015}, and a qualitative interpretation in terms of the analytically obtained SIM is given in \cite{Li.2016}. 
It is shown both numerically and experimentally that the periodic response with two symmetric impacts per period ceases to exist at the optimum design point (in terms of minimal vibration level of the primary oscillator) giving rise to Strongly Modulated Response \cite{Li.2016b}.
\\
The above mentioned analytical studies are all limited to host structures that can be accurately modeled as spring-mass-oscillators.
The test rigs for the experimental studies were designed accordingly, involving a stiff, bulky body suspended by a flexible, light spring.
In this case, it is appropriate to assume that the impacts cause immediate dissipation in the form of inelastic behavior within the contact region, which can be modeled in terms of a coefficient of restitution.
The coefficient of restitution has to be set based on experience or should be determined experimentally.
The picture changes if one attaches the impact absorber to a flexible host structure \cite{Theurich.2019}.
Then the impacts transfer energy from the resonant mode to high-frequency modes, where the energy is dissipated on faster time scales.
It has been demonstrated that this form of energy transfer is the main reason for the absorber's ability to mitigate the vibrations of flexible structures, for systems under either impulsive excitation \cite{AlShudeifat.2013,Luo.2014,AlShudeifat.2015} or harmonic loading \cite{Theurich.2019}.
From a technical perspective, it is interesting to design the contact so that inelastic behavior within the contact region is avoided as this is inevitably associated with damage.
The purpose of this work is to develop an approach to the design of impact absorbers for mitigating resonances of \emph{flexible host structures}.
To obtain a quantitatively accurate prediction of the optimum design, the energy transfer between the resonant mode and the high-frequency modes must be explicitly modeled.
This will be achieved by an appropriate resolution of the elasto-dynamics and the contact mechanics.
In \sref{semi}, a semi-analytical approach for this problem is presented.
In \sref{valid}, the quality and validity of the approach is assessed by comparison against direct numerical simulation.
This article ends with a synopsis of the main findings and certain conclusions in \sref{concl}.

\section{Development of the semi-analytical approach\label{sec:semi}}
\subsection{Problem setting\label{sec:problem}}
We consider a flexible, linear host structure with an attached impact absorber described by the system of equations of motion,
\begin{eqnarray}
	\bs M \ddot{\bs q}_\mrm s + {\bs D}\dot{\bs q}_\mrm s + \bs K \bs q_\mrm s + \bs w f_\mrm c =& \Re\lbrace \bs F_\mrm{ex}\ee^{\ii\Omega t}\rbrace\fk \label{eq:eom1} \\
	\ma \ddot{q}_\mrm a - f_\mrm c =& 0\fp\label{eq:eom2}
\end{eqnarray}
The host structure is described in terms of the vector of generalized coordinates $\bs q_\mrm s\in \mathbb R^{n\times 1}$, the symmetric and positive definite mass, damping and stiffness matrices, $\bs M, \bs D, \bs K$, respectively.
Over-dot denotes differentiation with respect to time $t$ and $\Re\lbrace.\rbrace$ denotes the real part of a complex vector.
The host structure is subjected to harmonic forcing with angular frequency $\Omega$ and complex amplitude vector $\bs F_\mrm{ex}\in\mathbb C^{n\times 1}$.
The kinematics of the impact absorber with mass $\ma$ is assumed to remain one-dimensional, so that it can be described by a single coordinate $q_{\mrm a}$.
Consequently, the interaction between the absorber and the host structure is modeled by a scalar contact force $f_{\mrm c}$. The force direction is described in \eref{eom1} by the vector $\bs w\in\mathbb R^{n\times 1}$.
The contact force, $f_{\mathrm c}$, is related to the relative displacement between the impact absorber and the point of attachment to the structure, $\delta = \bs w\tra{\mm q}_{\mrm s}-q_{\mrm a}$, via a \emph{contact law}, as discussed later.

\subsection{Central assumptions and overview of the semi-analytical approach\label{sec:assumptions}}
As described in \sref{intro}, it is an established hypothesis that the optimum design is reached where the (almost) periodic resonant response with two symmetric impacts per period ceases to exist, giving rise to a Strongly Modulated Response (\fref{schematicEfficacyCurve}).
It is therefore sufficient to limit the analysis to the periodic regime with two symmetric impacts per period, which is an important simplification.
To this end, we consider the case where the excitation frequency is near a well-separated natural frequency, the host structure is lightly damped and the absorber is light (i.e., the mass $\ma$ in \eref{eom2} is much smaller compared to the mass of the host structure).
Under these conditions, the dynamics of the host structure will be dominated by the fundamental harmonic of the resonant mode.
An according truncation of \eref{eom1} is carried out in \ssref{truncation}.
For the case of two symmetric impacts per period, the response of the impact absorber, ${q}_\mrm a$, resembles a sawtooth function.
The amplitude and phase of the sawtooth function are related to the amplitude of the host structure and the clearance via an effective modal coefficient of restitution, \cf \ssref{sawtooth}.
Here, the associated Slow Invariant Manifold is expressed as well.
To complete the governing system of equations, one needs to determine the modal coefficient of restitution and the contact force pulse, i.e. the time series of the interaction force between the host structure and the impact absorber.
Here, analytical half space theory is combined with a numerical short-time simulation of a single impact event in order to properly account for the energy transfer between the modes of the host structure that is inflicted by the impact pulses (\ssref{contact}).
The contact simulation yields both the contact force pulse and the modal coefficient of restitution (\ssref{prediction}).
On the one hand, the numerical simulation enables high accuracy and accounting for realistic boundary conditions.
On the other hand, this step renders the overall approach \emph{semi}-analytical.
Subsequently, we derive an equation governing the frequency response (\ssref{frequencyresponse}) and derive a closed-form expression of the resonant response (\ssref{closedform}).
\\
We remark that the theory in \ssref{truncation} and \ref{sec:sawtooth} could also be developed using the method of multiple scales, as in \cite{Gendelman.2012} and follow-up work.
Instead, we use an ad-hoc approach based on, among others, Harmonic Balance.
We are convinced that our approach can be much easier extended to host structures subjected to generic nonlinearities (\eg sophisticated models of dry friction damping).

\subsection{Truncation of the host structure dynamics to the fundamental harmonic of the resonant mode\label{sec:truncation}}
Following the assumptions stated in \ssref{assumptions}, the dynamics of the host structure is approximated by the fundamental harmonic of the resonant mode,
\begin{equation} \label{eq:resmode}
\bs \q_\mrm s \approx \Re\lbrace \bs \varphi  \amp \ee^{\ii\psiE}\ee^{\ii\Omega t}\rbrace\fp
\end{equation}
This assumes that the dynamics possesses a (fast) frequency close to the frequency of the resonant structural mode, and other (fast) frequencies (such as the frequencies of off-resonant modes) are assumed not to be present in the response. Herein, $\amp>0$ is the modal amplitude, and $\psiE$ is the phase lag to the excitation.
We denote by $\bs \varphi\in\mathbb R^{n\times 1}$ the mass-normalized deflection shape of the corresponding resonant mode of the underlying conservative linear system.
Substituting \eref{resmode} into \eref{eom1}, projecting onto $\bs \varphi$ and averaging over one period of oscillation yields the complex-valued algebraic equation
\begin{equation} \label{eq:averaged}
\left(-\Omega^2 + 2\D\omega\ii\Omega + \omega^2\right)a+ \phic \underbrace{\frac{1}{\pi} \int\limits_{0}^{2\pi} f_\mrm c \ee^{-\ii\left(\Omega t+\psiE\right)} \dd\Omega t}_{F_{\mrm c}\ee^{-\ii\psiE}} = {\bs \varphi}^\mrm T\bs F_\mrm{ex}\ee^{-\ii\psiE}\fp
\end{equation}
Herein, $\omega$ and $D$ are the modal angular frequency and modal damping ratio of the resonant mode, respectively.
The abbreviation $\phic = \bs \varphi^\mrm T \bs w$ is to denote the value of the mass-normalized modal deflection shape at the contact location.
Moreover, $F_{\mrm c}$ is the complex fundamental Fourier coefficient of the contact force $f_{\mrm c}$.
The real part of $F_{\mrm c}\ee^{-\ii\psiE}$ describes the inertia effects due to the intermittently attached absorber, and the imaginary part corresponds to the damping associated with the impacts.

\subsection{Sawtooth dynamics of the impact absorber and the Slow Invariant Manifold\label{sec:sawtooth}}
\begin{figure}[t!]
	\centering
	\includegraphics[width=0.75\textwidth]{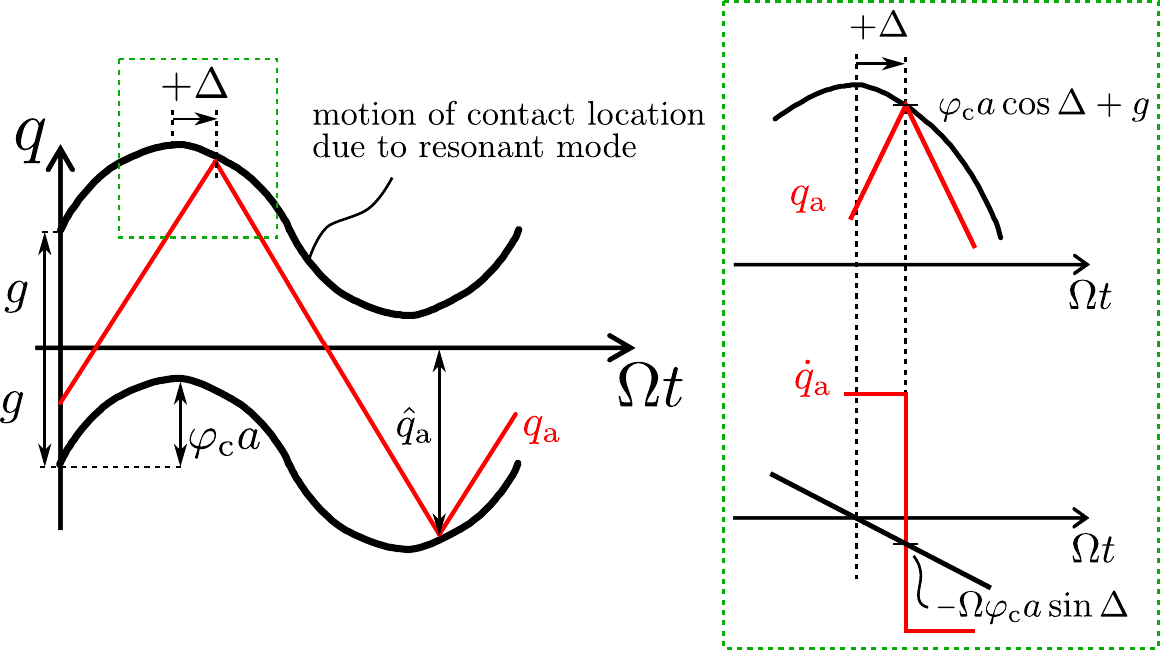}
	\caption{Illustration of the absorber and cavity motion for the case of two symmetric impacts per period.}
	\label{fig:absincav}
\end{figure}
%
The light absorber merely responds to the imposed harmonic movement of the host structure.
At the contact location, the host structure's elastic displacement is $\mm w\tra \mm q_{\mrm s}$.
Without loss of generality, we assume that the absorber displacement, $q_{\mrm a}$, is measured from the center of the cavity, so that the distance to either wall is $\gap>0$.
Hence, the cavity walls move as $\mm w\tra \mm q_{\mrm s}\pm \gap$.
Recall that two symmetric impacts per excitation period are considered in order to simplify the dynamical analysis (clearly, other response regimes exist but our approach focuses on this basic assumption).
On the (long) time scale of the excitation period, the dynamics of the absorber hence takes the form of a sawtooth function (\fref{absincav}).
In what follows, $\qahat$ denotes the amplitude of the absorber and $\psiA$ denotes the phase lag between the absorber and the host structure (this will be referred to as the contact phase angle).
One deduces that
\begin{equation}\label{eq:cosdel}
\cos\psiA = \frac{\qahat - \gap}{\phic a}\fp
\end{equation}
Here and in the following, we require, without loss of generality, that the contact force direction is defined in such a way that $\phic>0$.
\\
During each impact, energy is transferred (scattered) to elastic modes within the flexible host structure.
To model this process, an accurate spatial and temporal resolution of the impact event is necessary, as described in \ssref{contact} and \ref{sec:prediction}.
On the long time scale of the resonant mode and the rigid-body motion of the absorber, the impact event can be properly described by a finite force integral at a certain time instant and a coefficient of restitution $0\leq\cor<1$.
In this context, the coefficient of restitution is defined as the ratio between the relative velocities after and before the impact,
\begin{equation}\label{eq:cordef}
\cor = \frac{\left|\dot q_{\mathrm a}\right| - \phic a\Omega \sin\psiA}{\left|\dot q_{\mathrm a}\right| + \phic a\Omega \sin\psiA}\fp
\end{equation}
Note that the magnitude of the absorber velocity is the same before and after each impact, $\left|\dot q_{\mathrm a}\right| = 2\qahat/(\pi/\Omega)$.
In full accordance with the described time scale separation and the modal truncation, the velocity induced only by the resonant mode is considered in \eref{cordef}, as opposed to the physical velocity at the absorber location (which results from the superposition of all modal responses).
Consequently, we refer to $\cor$, defined in \eref{cordef}, as \emph{modal coefficient of restitution}.
For a coefficient of restitution $\cor<1$, \eref{cordef} states that the magnitude of the relative velocities is decreased by the impact, which corresponds to a loss of momentum, and thus mechanical energy, on the long time scale.
The mechanical energy is transferred irreversibly from the host structure's resonant mode to off-resonant modes.
\\
One can bring \eref{cordef} into the form
\ea{
\sin\psiA =\rho\frac{\qahat}{\phic a}\fk \quad \text{with}\,\, \rho =  \frac{2}{\pi}\frac{1-\cor}{1+\cor}\fp \label{eq:sindel}
}
Taking the square on both sides of \erefs{cosdel} and \erefo{sindel} and making use of the trigonometric identity, $\sin^2\psiA + \cos^2\psiA = 1$, one obtains a relation between the amplitude of the absorber, the amplitude of the host at the contact location (as induced by the resonant mode), the clearance and the coefficient of restitution (implicitly contained in the auxiliary quantity $\rho$):
\ea{
\left(\rho \qahat\right)^2 + \left(\qahat-\gap\right)^2 = \left(\phic\amp\right)^2\fp
}
This represents the Slow Invariant Manifold (SIM) of the resonant dynamics. Note that since only the movement induced by the resonant mode is considered, the expression of the SIM is identical to that presented by Gendelman \cite{Gendelman.2012} and studies based on that work.
Consequently, the existence and stability result obtained in \cite{Gendelman.2012} can be adopted.
It will later turn out to be convenient to parameterize the results in terms of the contact phase $\psiA$.
A range $\psiA_{\min}\leq \psiA\leq \psiA_{\max}$ can be determined, for which the stable branch of two symmetric impacts per period exists:
\ea{
\psiA_{\min} = \operatorname{atan}\rho\fk \qquad \psiA_{\max} = \operatorname{acos}\left(\frac{-\rho^2}{1+\rho^2}\right)\fp \label{eq:psiArange}
}
The lower bound, $\psiA_{\min}$, corresponds to the case where the amplitudes go to infinity, and it can be obtained by dividing both sides of \erefs{cosdel} and \erefo{sindel} and taking the corresponding limit.
The upper bound, $\psiA_{\max}$, corresponds to the smallest amplitude, $a$, of the host structure for which periodic oscillations exist (in essence, this corresponds to the fold point of the SIM).
For lower amplitudes than that, the periodic response ceases to exist, giving rise to a Strongly Modulated Response.
In the limit of $\cor\to 1$, $\psiA_{\max}\to\pi/2$, while $\psiA_{\max}\gtrsim\pi/2$ for $\cor<1$; \eg for $\cor=0.5$, $\psiA_{\max}$ is about $3\%$ larger than $\pi/2$.
It must be remarked that the branch of two symmetric impacts per period loses stability beyond a certain amplitude \cite{Gendelman.2012} giving rise to resonant responses with two asymmetric or more impacts per period.
\\
Finally, we eliminate $\qahat$ from \erefs{cosdel} and \erefo{sindel}, yielding
\ea{
\frac{\gap}{\phic \amp}=\frac 1\rho \sin\psiA-\cos\psiA\fk
\label{eq:apsir}
}
where we recall that $\rho>0$ is related to $\cor$ according to \eref{sindel}.
The relation in \eref{apsir} is illustrated in \fref{apsir} for different coefficients of restitution.
For any value of $\psiA$ one obtains a unique value of the normalized amplitude $\amp\phic/\gap$.
In fact, the relation is bijective in the valid range of amplitudes.
This will be exploited in \ssref{frequencyresponse} and \ref{sec:closedform}, where the analysis is carried out with respect to $\psiA$ rather than $\amp$.
\begin{figure}[t!]
	\centering
	\includegraphics[width=0.7\textwidth]{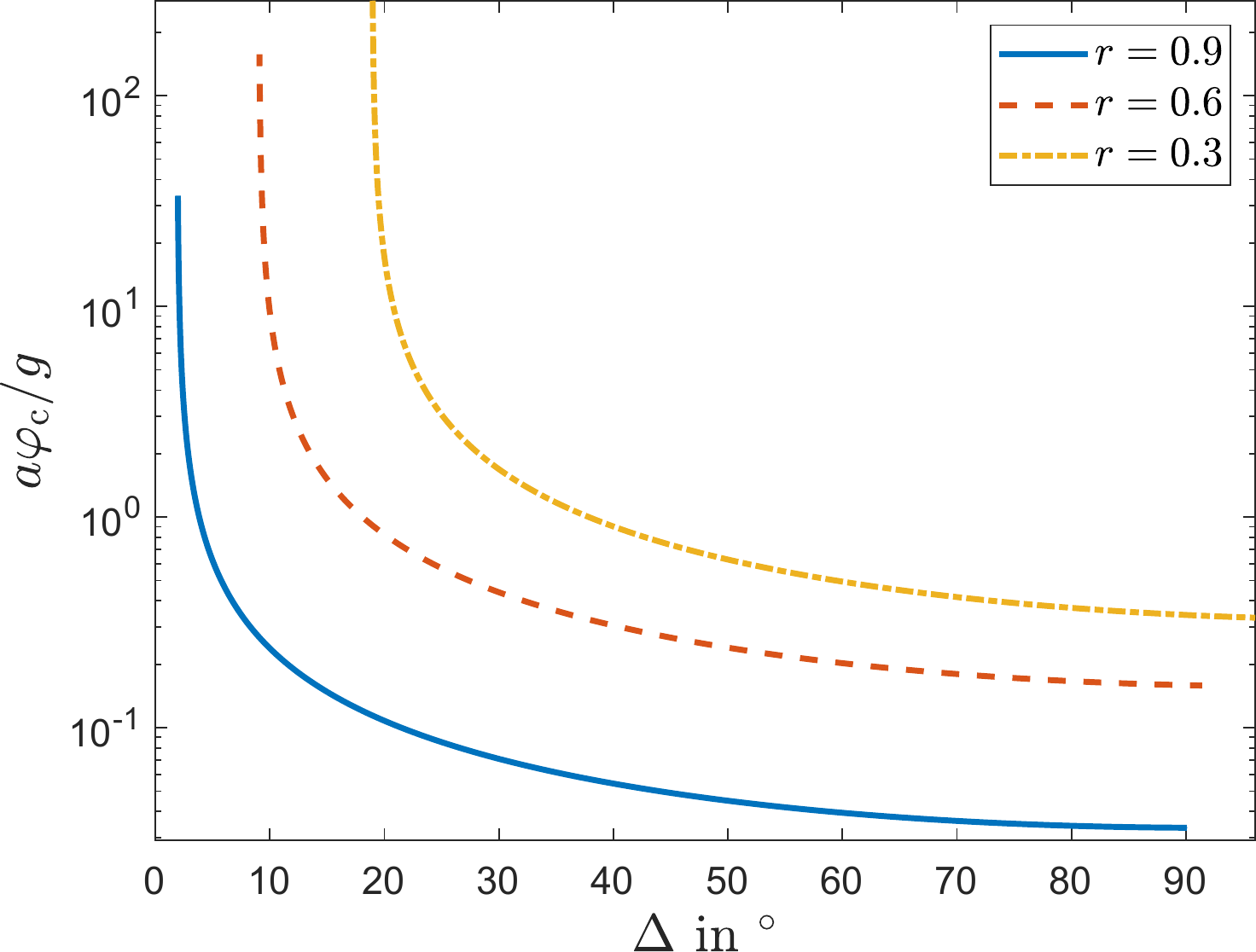}
	\caption{Relation between the amplitude of the host structure at the contact location, normalized by the clearance, and the contact phase angle for different coefficients of restitution.}
	\label{fig:apsir}
\end{figure}
\\
An alternative derivation of relation \ref{eq:apsir} by means of the method of Non-Smooth Temporal Transformation (NSTT) \cite{Pilipchuk.2015} is presented in Appendix A.

\subsection{Expression and averaging of the contact force pulse\label{sec:contact}}
To determine the frequency response from \eref{averaged}, we need to express the contact force $f_\mrm c$ and determine its fundamental Fourier coefficient $F_{\mrm c}$.
Previous studies modeled the contact using Newton's impact law and relied on an empirically determined coefficient of restitution.
Here, we intend to achieve a more predictive design approach by properly resolving the contact mechanics and elasto-dynamics, during an impact event, in time and in space.
In this particular aspect of our work, we assume that the contact between the host structure and the absorber can be adequately described using the Hertzian theory \cite{hert1881,john1989}.
This implies the assumption that the contact area remains small compared to the dimensions of the underlying bodies.
Moreover, Hertzian theory implies that the material behavior in the contact region remains linear-elastic.
This is reasonable given that inelastic behavior is not desired as it is inevitably associated with damage and wear within the contact region.
The validity of Hertzian theory for such problem settings is well supported by experiments, see \eg \cite{Seifried.2010}.
It should be remarked that Newton's impact law was used in previous work (\eg \cite{Theurich.2019}) instead. While the Hertzian model leads to a finite force pulse, Newton's impact law commonly leads to infinitesimally short force pulses corresponding to Dirac-delta distributions. In accordance with the Fourier transform of this distribution, Newtonian impacts generally scatter energy to arbitrarily high frequencies. In numerical practice, this may lead to lacking modal convergence. For these reasons, Newton's impact law should not be used in the current problem setting and the Hertzian model is much more appropriate.
Note that Hertzian theory only describes the normal contact interactions, while the potential effects of dry friction in the tangential contact plane are neglected. 
\\
Applying Hertzian theory to the contact between the absorber and the host structure leads to the following expression of the contact force $f_{\mrm c}$:
\ea{
f_{\mrm c} = k_{\mrm H}
 \begin{cases}
-\left|\delta+\gap\right|^{3/2} & \delta<-\gap \\
0 & -\gap\leq\delta\leq\gap \\
\left|\delta-\gap\right|^{3/2} & \delta>\gap
\end{cases}
\label{eq:hertz}
}
with $\delta= \bs w\tra\mm q_{\mrm s}-q_{\mrm a}$.
In this work, we consider the common case of a spherical absorber contact interface, absorber radius $R_{\mrm a}$, and isotropic material behavior within the contact region, with Young's modulus $E_{\mrm a}$, $E_{\mrm s}$, and Poisson ratio $\nu_{\mrm a}$, $\nu_{\mrm s}$ of absorber and host structure, respectively.
The Hertzian spring constant $k_{\mrm H}$ is then given by
\ea{
k_{\mathrm H} = \frac 4 3 \sqrt{R_{\mrm a}} \left( \frac{1-\nu_{\mrm a}^2}{E_{\mrm a}} + \frac{1-\nu_{\mrm s}^2}{E_{\mrm s}} \right)^{-1}\fp
\label{eq:hertzspring}
}
%
For the scenario of a sphere, of mass $m_{\mrm a}$, impacting on an elastic half space, with velocity $v_{\mrm c}>0$, an analytical approximation of the contact force pulse has been established by Hunter \cite{Hunter.1957} and Reed \cite{Reed.1985}.
In this case, the force pulse is given by $\fc\sin^{3/2}(\pi \overline t/\tc)$ for the duration of the contact event $0\leq \overline t\leq \tc$, where time $\overline t$ counts from the beginning of the impact.
The maximum contact force, $\fc$, and contact duration, $\tc$, are expressed as
\ea{
\tc &=& 2.94 \frac{\alpha_0\tilde\alpha}{v_{\mrm c}}\tilde T_{\mrm c}\fk \label{eq:tc}\\
\fc &=& k_{\mrm H}\left(\alpha_0\tilde\alpha\right)^{\frac 32}\fk \label{eq:fc} \\
\text{with}\,\, \alpha_0 &=& \left(\frac 5 4 \frac{m_{\mathrm a}}{k_{\mrm H}} v_{\mrm c}^2\right)^{\frac 2 5} \fp \label{eq:alpha0}
}
In contrast to the original Hunter-Reed approximation \cite{Hunter.1957,Reed.1985}, we have introduced here the correction factors $\tilde\alpha$ and $\tilde T_{\mrm c}$ for the maximum compression and the contact duration, respectively.
These are needed to take into account the finite dimensions and boundary conditions of the host structure.
Moreover, we believe that the augmented Hunter-Reed approximation in \erefs{tc}-\erefo{fc} will be useful for non-Hertzian (yet elastic) contacts as well.
The correction factors are numerically predicted as described later in \ssref{prediction}. 
For illustration, an example of the original and the corrected force pulse is depicted together with the numerical reference in \fref{correctedfc}.
\begin{figure}[b!]
	\centering
	\includegraphics[width=0.5\textwidth]{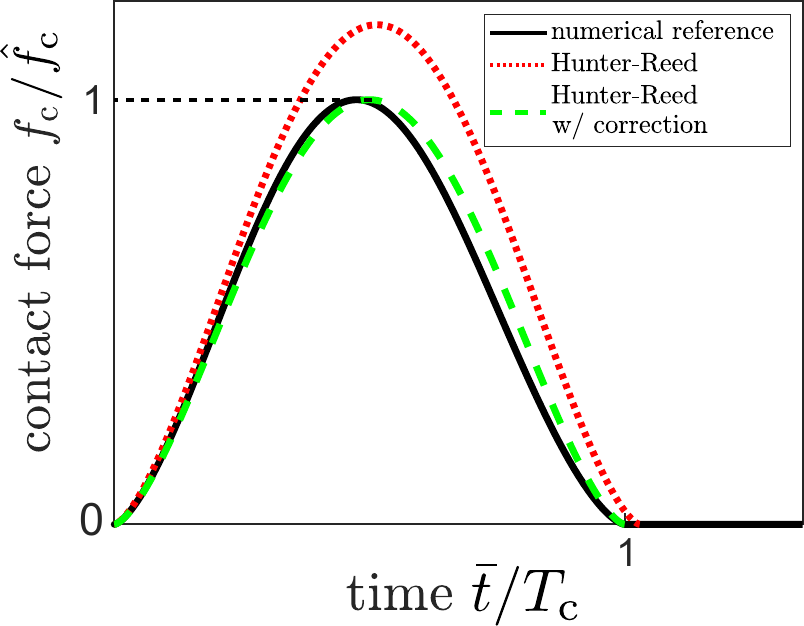}
	\caption{Correction of the contact duration and maximum contact force via the impact event simulation.}
	\label{fig:correctedfc}
\end{figure}
\\
We now employ the augmented Hunter-Reed approximation to determine the complex Fourier coefficient $F_{\mrm c}$ of the contact force $f_{\mrm c}$ in \eref{averaged}.
In accordance with \fref{absincav}, the first impact occurs at $\Omega t + \psiE = \psiA$.
During this impact, we have $\delta<-\gap$, so the contact force is actually negative, $f_{\mrm c}<0$, \cf \eref{hertz}.
The next impact occurs at $\Omega t + \psiE = \psiA + \pi$, where the contact force has the opposite sign.
Due to the assumed symmetry of the two impacts within a period, it is sufficient to take the integral in \eref{averaged} only over the first impact and multiply the result by two to include the effects of both impacts.
In accordance with the previously described time scale separation, the argument $\Omega t + \psiE$ within \eref{averaged} is approximated as being constant for the short duration of the impact.
Hence the term $\ee^{-\ii\psiA}$ can be moved outside the integral.
With these considerations, one obtains the approximation
\ea{
F_{\mrm c}\ee^{-\ii\psiE}
= -\frac 2 \pi \ee^{-\ii\psiA} \Omega\tc \fc \int_{0}^{1} \sin ^{\frac 3 2} \left(\pi\tau\right)\dd \tau\fk \label{eq:Fc}
}
where $\tau = \overline t/\tc$.
At this point we can expand the term $\tc\fc$ in \eref{Fc} by substituting \erefs{tc}-\erefo{alpha0},
\ea{
\tc\fc
&=& 2.94  \tilde T_{\mrm c} k_{\mrm H} \frac{\left(\alpha_0\tilde\alpha\right)^{\frac 5 2}}{v_{\mrm c}} \nonumber\\
&=& 2.94 \cdot \frac 5 4  \tilde T_{\mrm c} \tilde\alpha^{\frac 5 2} m_{\mrm a} v_{\mrm c}\fp \label{eq:tcfc}
}
%
At the considered first impact (\fref{absincav}), the pre-impact relative velocity, $v_{\mrm c}$, is given by
\ea{
v_{\mrm c}
&=& \Omega \left[\phic a \sin\psiA + \frac 2 \pi \left(\phic a \cos\psiA + \gap\right)\right] \nonumber\\
&=& \Omega\phic a \left[ \sin\psiA + \frac 2 \pi \left(\cos\psiA + \frac{\gap}{\phic a}\right)\right] \nonumber\\
&=& \Omega\phic a \left[ 1 + \frac2 \pi \frac 1\rho \right] \sin\psiA \nonumber\\
&=& \Omega\phic a \frac{2}{1-\cor} \sin\psiA\fk \label{eq:vc}
}
where we have substituted \eref{apsir} and the definition of $\rho$ from \eref{sindel}.
Note that the Hunter-Reed approximation requires that $v_{\mrm c}>0$, and recall that we assumed that $\phic>0$ and $a>0$ without loss of generality.
Also note that $0\leq\cor\leq 1$ and $\sin\psiA>0$ for the range of $\psiA$ defined in \eref{psiArange}.
To ensure that $v_{\mrm c}>0$ in \eref{vc}, we require that $\Omega>0$ henceforth, which will be important for the derivation of a closed-form solution, as outlined below.

\subsection{Frequency response equation\label{sec:frequencyresponse}}
Substituting \erefs{Fc}, \erefo{tcfc} and \erefo{vc} into \eref{averaged}, we obtain
\ea{
\left(-\Omega^2 + 2\D\omega\ii\Omega + \omega^2\right)a
-
~
\underbrace{\frac 2 \pi \cdot 2.94 \cdot \frac 5 4  \tilde T_{\mrm c} \tilde\alpha^{\frac 5 2} \int_{0}^{1} \sin ^{\frac 3 2} \left(\pi\tau\right)\dd \tau}_{\Gamma}
~
\underbrace{m_{\mrm a} \phic^2}_{\mmr}
~
a \ee^{-\ii\psiA} \Omega^2
~
\frac{2}{1-\cor} \sin\psiA
=
\fex\ee^{-\ii\psiE}\fk
}
where we used the abbreviation $\fex = {\bs \varphi}^\mrm T\bs F_\mrm{ex}$ and $\mmr$ is the modal mass ratio.
Finally, introducing the frequency ratio $\eta = \Omega/\omega$ and eliminating $a$ via \eref{apsir} yields
\ea{
1 - \eta^2\left(1+\frac{2\mmr\Gamma}{1-\cor}\sin\psiA\ee^{-\ii\psiA}\right) + 2 D\eta \ii &=& \frac{\fex }{\omega^2 a}\ee^{-\ii\psiE} \fk \\
1 - \eta^2\left(1+\frac{2\mmr\Gamma}{1-\cor}\sin\psiA\ee^{-\ii\psiA}\right) + 2 D\eta \ii &=& \frac{\fex \phic}{\omega^2 \gap}\left(\frac \pi 2 \frac{1+\cor}{1-\cor} \sin\psiA - \cos\psiA\right)\ee^{-\ii\psiE} \fp \label{eq:frequencyresponse}
}
\eref{frequencyresponse} is a complex algebraic equation in the three real unknowns $\eta$, $\psiA$ and $\psiE$, and thus governs the \emph{frequency response} of the host structure with the absorber, in the regime of two symmetric impacts per period. From \eref{frequencyresponse} one can follow that the design problem reduces to four dimensionless parameters:
\begin{itemize}
	\item $D$: damping ratio of the resonant mode
	\item $\mmr\Gamma$: modal mass ratio multiplied with contact force pulse properties
	\item $\cor$: modal coefficient of restitution
	\item $\frac{\fex\phic}{\omega^2\gap}$: static deflection of the resonant mode at the contact location, normalized by the clearance
\end{itemize}
%

\subsection{Closed-form expression for the resonant response\label{sec:closedform}}
By taking the magnitude squared on both sides of \eref{frequencyresponse} one can eliminate the phase lag $\psiE$ between the host structure and the excitation:
\ea{
\left(1-\eta^2\left(1+\frac{2\mmr\Gamma}{1-\cor}\sd \cd\right)\right)^2+\eta^2\left(2\D+\eta\frac{2\mmr\Gamma}{1-\cor}\sin^2\psiA\right)^2 =\nonumber\\
	\left(\frac{\fex\phic}{\gap\omega^2}\right)^2 \left(\frac{\pi}{2}\frac{1+\cor}{1-\cor}\sd-\cd\right)^2\fp\label{eq:ampfreq}
}
\eref{ampfreq} governs the relation between the normalized excitation frequency $\eta$ and the contact phase $\psiA$.
Recall that for any value of $\psiA$, one obtains a unique amplitude $a$ via \eref{apsir}.
\begin{figure}[b!]
	\centering
	\includegraphics[width=0.55\textwidth]{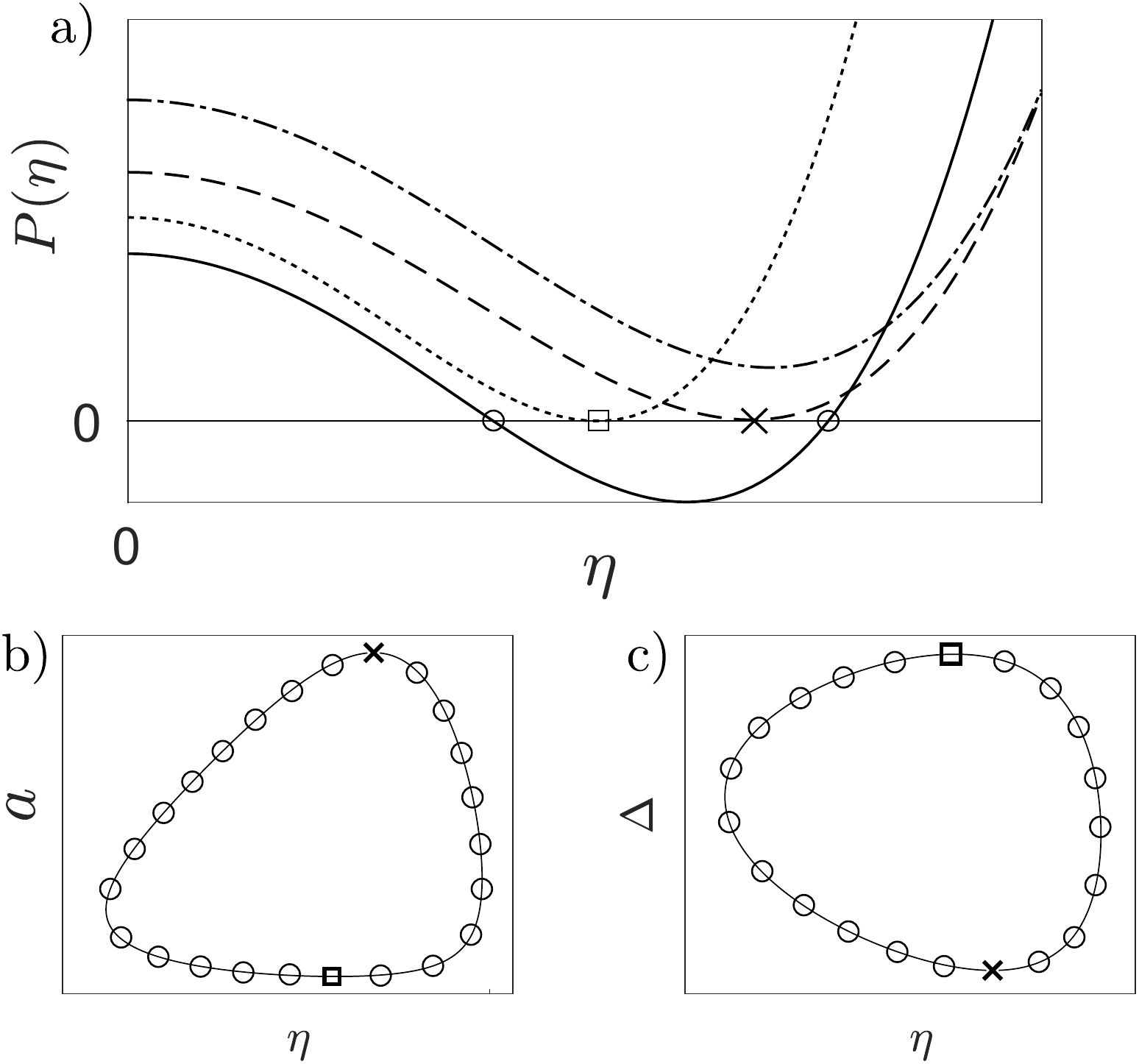}
	\caption{Study of the quartic polynomial in \eref{poly}: (a) The cases of zero, one and two positive real roots, (b) and (c) corresponding points on the amplitude-frequency curve and on the contact phase-frequency curve, respectively.}
	\label{fig:poly}
\end{figure}
%
Note that \eref{ampfreq} is a quartic polynomial equation in $\eta$,
\ea{
P(\eta)= A\eta^4 + B\eta^3 + C\eta^2 +E = 0\fk \label{eq:poly}
}
with coefficients
\ea{
A &=&  \left(1+\frac{2\mmr\Gamma}{1-\cor}\sd \cd\right)^2+\left(\frac{2\mmr\Gamma}{1-\cor}\sin^2\psiA\right)^2>0\fk\\
B &=& \frac{4\mmr\Gamma}{1-\cor}2\D\sin^2\psiA>0\fk\\
C &=& -2\left(1+\frac{2\mmr\Gamma}{1-\cor}\sd\cd\right)+\left(2\D\right)^2<0 \quad\text{ if } |\D|<\frac{1}{\sqrt{2}}\fk\\
E &=& 1-\left(\frac{\fex\phic}{\gap\omega^2}\right)^2 \left(\frac{\pi}{2}\frac{1+\cor}{1-\cor}\sd-\cd\right)^2\fp
}
The polynomial $P(\eta)$ is schematically depicted in \fref{poly}a.
Recall that we required $\Omega>0$ and thus we only need to consider $\eta>0$ (\cf \ssref{contact}).
Taking into account the signs of the individual coefficients, $P(\eta)$ can have zero, one or two positive real roots, as indicated in \fref{poly}a.
The case of one (double) root corresponds to an extremum of the $\psiA$-$\eta$ curve, and thus $\amp$-$\eta$ curve (\fref{poly}b-c).
The extremum corresponds either to a maximum or a minimum of the amplitude-frequency curve.
A minimum occurs if the amplitude-frequency curve takes the form of an isolated branch.
We are interested in the resonant response, \ie, in the maximum amplitude.
The idea will be to look for double roots and distinguish between minimum and maximum amplitudes later.
\\
For a double root, we must have that both conditions $P=0$ and $\partial P/\partial \eta = 0$ are satisfied for the same value of $\eta$.
The condition $\partial P/\partial \eta = 0$ leads to:
\ea{
\partial P/\partial \eta = 4A\eta^3 + 3B\eta^2 + 2C\eta \stackrel{!}{=} 0 \\
\Rightarrow \, \eta_0 = 0\fk \eta_{1,2} = \frac{3B}{8A}\left(-1\pm\sqrt{\underbrace{1-\frac{32AC}{9B^2}}_{Q}}\right)\fp \label{eq:eta1}
}
Note that $Q>0$ is ensured due to the signs of the coefficients $A$, $B$ and $C$.
Of the three roots, only $\eta_1$ (with the plus sign) is positive and is thus considered further.
Substituting $\eta_1$ into \eref{poly} yields, after some algebraic manipulation,
\ea{
Q^{3/2}-\frac{3}{2}Q-\frac{3}{8}(Q-1)^2+\frac{1}{2}+\frac{512A^3E}{27B^4} = 0 \label{eq:critresp}\\
\Rightarrow \left(\frac{\fex\phic}{\gap\omega^2}\right)^2 = \frac{\frac{27B^4}{512A^3}\left(Q^{3/2}-\frac{3}{2}Q-\frac{3}{8}(Q-1)^2+\frac{1}{2}\right)+1}{\left(\frac{\pi}{2}\frac{1+\cor}{1-\cor}\sd-\cd\right)^2}\fp \label{eq:critresp2}
}
Here, it should be remarked that $Q$ does not depend on $\fex\phic$.
For the design of the impact absorber, it is useful to normalize the amplitude, $\ares$, by the resonant amplitude without absorber,  $\aresnoabs = \fex/(2D\omega^2)$.
Similarly, we normalize the clearance, $\gap$, by the displacement at the absorber location for the resonant case without absorber, $\phic\aresnoabs$.
With these normalizations and considering \eref{apsir}, we can employ \eref{critresp2} to finally obtain:
\ea{
\frac{\ares}{\aresnoabs} &=& \frac{2D}{\sqrt{\frac{27B^4}{512A^3}\left(Q^{3/2}-\frac{3}{2}Q-\frac{3}{8}(Q-1)^2+\frac{1}{2}\right)+1}}\fk \label{eq:aresnorm} \\
\frac{\gap}{\phic\aresnoabs} &=& 2D \frac{\frac{\pi}{2}\frac{1+\cor}{1-\cor}\sd-\cd}{\sqrt{\frac{27B^4}{512A^3}\left(Q^{3/2}-\frac{3}{2}Q-\frac{3}{8}(Q-1)^2+\frac{1}{2}\right)+1}}\fp \label{eq:gapnorm}
}
\fref{effisola}a depicts $\ares/\aresnoabs$ vs. the normalized clearance $\gap/(\phic\aresnoabs)$.
This curve is obtained as follows:
First, the dimensionless parameters $D$, $\cor$ and $\mmr\Gamma$ are assigned set values (the specific values for \fref{effisola} are given in the caption).
Then the upper and lower bounds of $\psiA$ are determined as defined in \eref{psiArange} for the given $\cor$.
Finally, \erefs{aresnorm}-\erefo{gapnorm} are evaluated numerically for a sequence of discrete values of $\psiA$ within these bounds.
\\
For sufficiently small clearance, there is only a single extremum of the amplitude, and this is a maximum (\fref{effisola}c).
For larger clearance, the amplitude-frequency curve takes the form of an isolated branch (\fref{effisola}b).
The dashed curve corresponds to the minimum and the solid line to the maximum amplitude.
In the following, the minimum amplitude is ignored so that only the range up to the turning point in \fref{effisola}a is considered. Hence, $\ares$ indeed corresponds to the \emph{resonant amplitude}.
At a certain level of the normalized clearance, the minimum and maximum of the amplitude-frequency curve coincide.
Here, the isolated branch degenerates to a point.
As discussed in \sref{intro}, this corresponds to the \emph{optimum design point}.
Beyond this normalized clearance, the periodic response ceases to exist, and the resonant response is strongly modulated.
\begin{figure}[t!]
	\centering
	\includegraphics[width=0.9\textwidth]{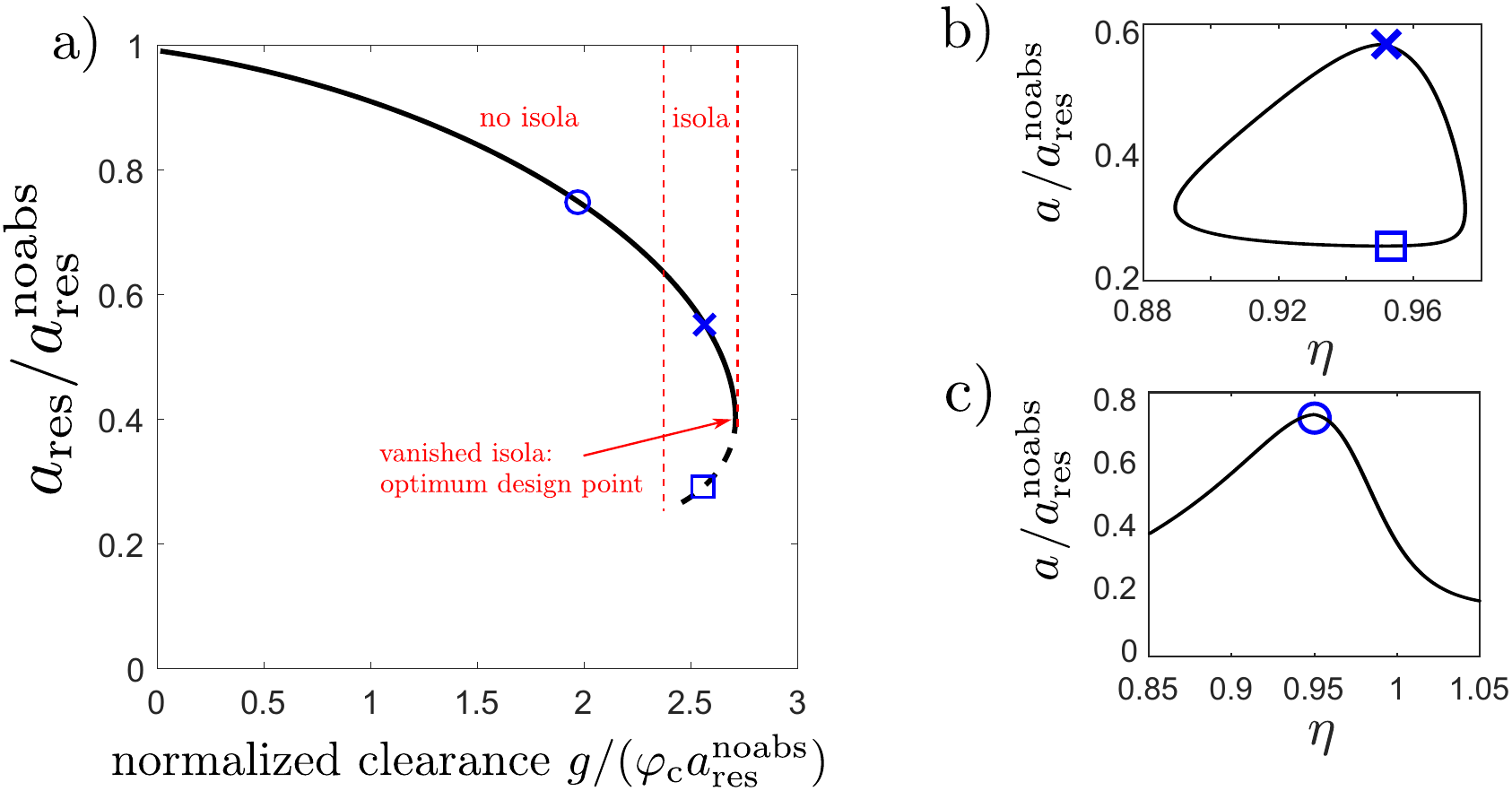}
	\caption{(a) normalized resonant response vs. normalized clearance and (b)-(c) amplitude-frequency curves for the case with and without isolated branch; parameters are  $\D=0.05, \mmr\Gamma=0.043, \cor=0.71$; the normalized clearance is set in (b) $\frac{\gap}{\phic\aresnoabs}=2.6$ and in (c) $\frac{\gap}{\phic\aresnoabs}=2.$}
	\label{fig:effisola}
\end{figure}

\subsection{Numerical prediction of the modal coefficient of restitution and the correction factors of the contact force pulse\label{sec:prediction}}
Recall that the assumed sawtooth dynamics of the absorber is described using the modal coefficient of restitution, $\cor$,
and that the averaged contact force pulse is described using the correction factors $\tilde\alpha$ and $\tilde T_{\mrm c}$.
In principle, these parameters could be determined experimentally.
To achieve a predictive design strategy, however, we propose to compute these parameters by a high-fidelity simulation of the impact event.
This requires an accurate spatial and temporal resolution of the elastic deformation of the contact region and the resulting elastic wave propagation (within the host structure).
\\
In the numerical example, we consider an Euler-Bernoulli beam hosting a spherical absorber.
Hence, the contact between the host structure and the absorber is modeled with a Hertzian spring as defined in \erefs{hertz} and \erefo{hertzspring}.
To resolve the elasto-dynamics of the host structure, we derive a model in the form of \eref{eom1} by truncating to a sufficiently large set of the lowest-frequency bending modes in accordance with the conventional Euler-Bernoulli beam theory.
The simulation is carried out using time step integration as described in \sref{valid}.
For more complicated configurations, we propose to use an appropriate finite element model.
\\
We propose to determine the parameters $\cor$, $\tilde\alpha$, and $\tilde T_{\mrm c}$ from a \emph{single representative impact event}.
For this event, the host structure has no initial deformation or velocity, ${\mm q}_{\mrm s}\left(0\right) = \mm 0 = \dot{\mm q}_{\mrm s}\left(0\right)$,
while the contact is closed at the start of the simulation, $q_{\mrm a}\left(0\right)=g$, and the absorber has an initial velocity $\dot q_{\mrm a}\left(0\right) = v_{\mrm c}>0$.
The simulation is run until time $T_{\mrm c}$, where the contact opens and the absorber has velocity $\dot q_{\mrm a}\leq 0$.
By recording the force in the Hertzian spring, one can directly determine the maximum, $\fc$ (\cf \fref{correctedfc}).
With this maximum, and the contact duration, $T_{\mrm c}$, it is possible to determine the correction factors $\tilde\alpha$ and $\tilde T_{\mrm c}$ using \erefs{tc}-\erefo{fc}.
To this end, we substitute the numerically obtained $\fc$ and $\tc$ on the left-hand side of \erefs{tc}-\erefo{fc} and solve for $\tilde\alpha$ and $\tilde T_{\mrm c}$.
The coefficient of restitution, $\cor$, with respect to the resonant mode, $\mm\varphi$, is obtained as
\ea{
\cor = \frac{\left|\dot q_{\mathrm a}\left(T_{\mrm c}\right) - \phic~\mm\varphi\tra\mm M\dot{\mm q}_{\mrm s}\left(T_{\mrm c}\right)\right|}{v_{\mrm c}}\fk \label{eq:corsim}
}
where $\mm\varphi\tra\mm M\dot{\mm q}_{\mrm s}$ extracts the current value of the modal velocity (of the resonant mode).
\\
Using the modal coefficient of restitution in the proposed way implies the assumption of an irreversible energy transfer from the resonant to the off-resonant modes.
One may object to this assumption that the off-resonant modes interact in a nonlinear way with the resonant mode during an impact, and that the dynamics of the off-resonant modes affect when and with what relative velocity the next impact(s) occur.
The off-resonant modes with much higher frequency can be expected to have decayed almost entirely between two subsequent impacts (half period of resonant mode).
Depending on their damping ratios, however, some of the lower-frequency modes will not have decayed as much.
One should stress, however, that this does not necessarily mean that a significant portion of energy accumulates in these modes from impact to impact.
This is because the impact does not generally occur at the same phase of the free oscillation.
Consequently, the energy contained in the resonant mode typically dominates the dynamics, and the interaction with the off-resonant modes does not invalidate the assumption of irreversible energy transfer.
However, the above described interaction with the off-resonant modes is largely responsible for the chaotic character of the dynamics in the almost periodic regime with on average two impacts per period.
\begin{figure}[b!]
	\centering
	\includegraphics[width=0.9\textwidth]{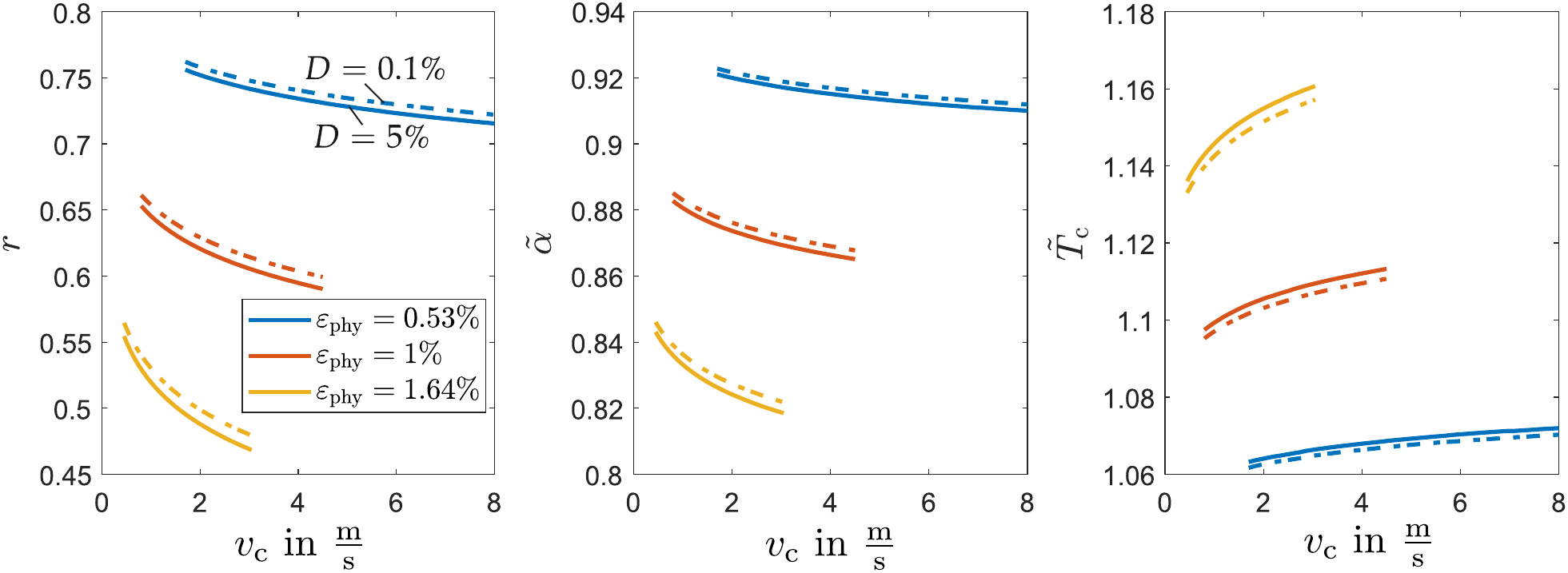}
	\caption{Dependence of modal coefficient of restitution, $\cor$, and correction factors of the contact force pulse,  $\tilde\alpha, \tilde T_{\mrm c}$ on the pre-impact velocity $\vc$ and the modal damping ratio $\D$ (set equal for all modes) for different mass ratios. The results are shown only in the interval of $\vc$ corresponding to the optimum design point, which depends on the respective $\pmr$ and the considered range of the damping ratio.}
	\label{fig:CoR_correctionFactors_vs_vc_D}
\end{figure}
\\
In \fref{CoR_correctionFactors_vs_vc_D}, the parameters $\cor$, $\tilde\alpha$, and $\tilde T_{\mrm c}$ are depicted as functions of the pre-impact velocity $v_{\mrm c}$ for different absorber masses, $\ma$, and for different modal damping ratios, $D$.
The absorber mass is indicated in terms of the physical mass ratio $\pmr={\ma}/{m_\mrm s}$ where $m_\mrm s$ is the mass of the host structure.
The results are shown for the numerical example of a cantilevered beam hosting a spherical impact absorber, as detailed in \sref{valid}.
The results reveal that the parameters $\cor$, $\tilde\alpha$, and $\tilde T_{\mrm c}$ have almost no explicit dependence on damping within the broad range between $D=0.1\%$ and $D=5\%$.
Moreover, the dependence on $\vc$ is relatively weak.
Note that in the periodic response regime, $\vc$ depends on the clearance, the amplitude and the frequency (\eref{vc}).
The value of $\vc$ at the optimum design point depends on the absorber mass and the damping ratio.
%
%
We propose to carry out the numerical simulation of the impact event just for a single representative value of $v_{\mrm c}$.
In order to increase accuracy near the optimum design point, we propose a two-step procedure:
(i) Perform an impact simulation for $\vc$ according to \eref{vc} with $\Omega = \omega$, $\amp = \aresnoabs$, $\gap = \phic \amp$, $\psiA = \pi/4$, and extract the parameters $\cor$, $\tilde\alpha$, and $\tilde T_{\mrm c}$;
(ii) determine the optimum design point as described in \ssref{closedform} and compute the corresponding value of $\vc$ at the optimum point;
and (iii) redo the impact simulation with this updated value of $\vc$.
Throughout the numerical studies in \sref{valid}, we found that such a single re-iteration is sufficient.
Thanks to the weak dependence on $\vc$, further iterations did not significantly change the results.
The proposed re-iteration is, however, important as it was found to improve the accuracy of the extracted parameters by up to $\approx 13\%$.
The more pronounced the vibration mitigation effect of the absorber is, the more important the proposed re-iteration becomes, since the amplitude at the optimum is considerably reduced from the initially assumed one in step (i).
\\
One should emphasize that the computational effort for the simulation of a single impact event is small compared to the simulation of the long-term forced vibration (involving many impacts).
Note, however, that the simulation generally has to be repeated if the properties of the host structure (mass and stiffness distribution, boundary conditions), or of the contact (location, effective stiffness, curvature radius) are modified.

\subsection{Comparison to the approach proposed in \cite{Gendelman.2012} and follow-up studies}
We remarked earlier that the expression of the Slow Invariant Manifold is identical to that obtained in \cite{Gendelman.2012}.
The assumptions of the time scale separation and the dominance of the fundamental harmonic of the host system's response are also in agreement with \cite{Gendelman.2012} and follow-up studies.
Since we furthermore propose to truncate to a single mode, the following question arises: In what way does our approach differ from that proposed in \cite{Gendelman.2015} for a single-degree-of-freedom oscillator?
We answer this question by noting that the important difference is that our point of departure is a flexible host structure described by \eref{eom1}, rather than a single spring-mass oscillator.
The modal truncation then leads to the notion of a \emph{modal coefficient of restitution}, \ie, a coefficient of restitution that considers only the contribution of the resonant mode to the pre- and post-impact velocities, as defined in \eref{cordef}.
Moreover, we propose to \emph{predict} this modal coefficient of restitution, by means of a detailed resolution of the contact mechanics and elasto-dynamics as described in \ssref{prediction}, rather than setting an empirically obtained and constant value.
In the numerical example, we use Hertzian theory to describe the contact, which is well-supported by experiments.
Our approach also leads us to an expression of the contact force pulse, which differs from that based on a pure Newtonian impact in \cite{Gendelman.2012} and follow-up work.
Comparing very carefully \eref{frequencyresponse} with the frequency response equation obtained in \cite{Gendelman.2015}, one can establish that the constant factor $2/\pi$ in \cite{Gendelman.2015} has to be replaced by the factor $\Gamma/(1+\cor)$ to arrive at the same expression related to the contact force pulse in this work.
Our factor, $\Gamma/(1+\cor)$, takes into account the shape of the contact force pulse, along with the correction factors and the predicted coefficient of restitution.
We have evaluated the factor, $\Gamma/(1+\cor)$, for some parameter combinations of our numerical example, and indeed obtained a value close to but not equal to $2/\pi$.
This suggests that it may be a valid strategy to adopt the theory in \cite{Gendelman.2012} and follow-up work, and merely replace the properties of the spring-mass oscillator by that of the resonant modal oscillator, and to substitute the predicted modal coefficient of restitution.
A further analysis of the theoretical differences is left for future work.

\section{Validation of the semi-analytical approach against numerical reference} \label{sec:valid}
\begin{figure}[b!]
	\centering
	\includegraphics[width=0.9\textwidth]{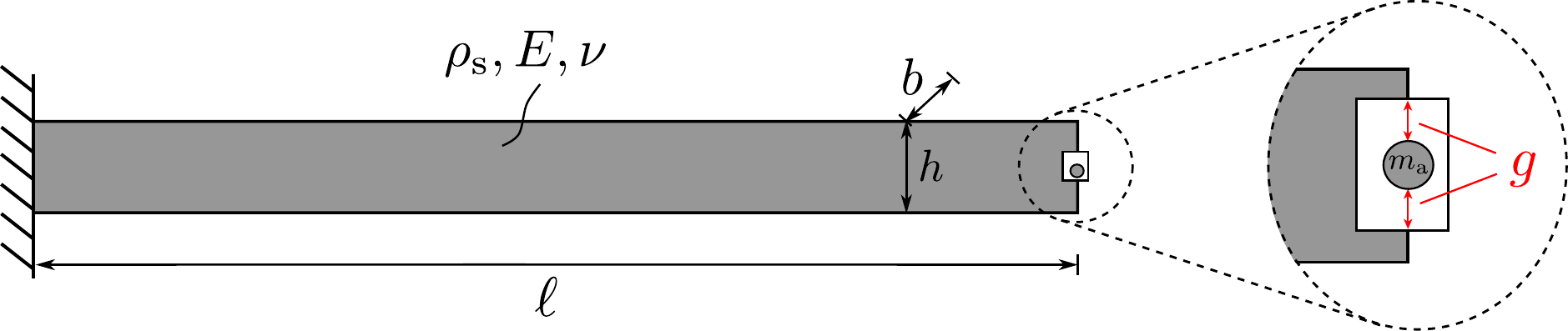}
	\caption{Benchmark model consisting of a cantilevered Euler-Bernoulli beam with spherical absorber at its tip.}
	\label{fig:beam}
\end{figure}
\begin{table}[b!]
	\centering
	\caption{Nominal parameters of the cantilevered Euler-Bernoulli beam.}
	\begin{tabular}{cccccccc}
		\hline
		Quantity & length $\ell$ & height $h$ & width $b$ & density $\rho_\mrm s$ & $E$ & $\nu$ & $\sigma_\mrm{b,W}$\\
		Value & 0.21 m & 0.01 m & 0.015 m & 7800 kg/m$^3$ & 210 GPa & 0.3 & 255 MPa\\
		\hline
	\end{tabular} \label{tab:model}
\end{table}
In order to assess the predictive capabilities of the developed semi-analytical approach, we compare the results against direct numerical simulation.
As host structure, we consider a cantilevered Euler-Bernoulli beam with uniform rectangular cross section. Its geometrical and material properties are given in \tref{model}, and it is schematically depicted in \fref{beam}.
A modal truncation to the lowest-frequency bending modes (without absorber) is carried out to obtain equations of motion in the form of \eref{eom1}.
Based on a convergence study, a truncation order of 12 was deemed sufficient.
The damping ratios of all modes are set equal to that of the resonant one, namely $D$.
A spherical absorber is placed at the beam's free end.
The absorber material is the same as that of the beam.
The contact interfaces on the beam side are considered as planar.
Thus, the radius of the spherical absorber, $R_{\mrm a}$, controls both the absorber mass, $m_{\mrm a}$, and equals to the effective contact radius within the Hertzian theory.
A concentrated harmonic force is applied at a length equal to $1/3\ell$ from the clamping in the beam's height direction.
The excitation frequency range near the primary resonance with the lowest-frequency bending mode is considered.
The excitation level is chosen such that the maximum bending stress in the resonant case without the absorber attached reaches the fatigue strength against bending, $\sigma_\mrm{b,W}$ (\tref{model}).
%
For numerical reference, the steady-state response to a stepped sine excitation with increasing frequency is computed.
Here, 20 equidistant levels of the excitation frequency ratio in the range $0.9\leq \eta\leq 1.07$ are considered.
Upon each frequency step, the initial values are adopted from the previous frequency level.
The simulation is run until the results stabilize with respect to a given tolerance of $1\%$.
In this numerical example, the particular challenges are the chaotic character of the response and the fact that the response may be strongly modulated.
To reach a stabilized mean amplitude, it was usually sufficient to simulate 300 excitation periods and extract the mean amplitude from the last 200 excitation periods.
To reach a stabilized maximum amplitude though, sometimes several thousands of excitation periods were needed (this was the case for Strongly Modulated Responses).
The third-order Bogacki-Shampine integration scheme (Matlab's $\operatorname{ode3}$) was used with a fixed time step.
Here, the time step was chosen such that the period of the retained highest-frequency mode was sampled at 15 points, which was found to yield a reasonable compromise between accuracy and computational cost.

\subsection{Numerical assessment of the basic assumptions of the semi-analytical approach} \label{sec:verify}
First, we analyze to what extent the basic assumptions of the semi-analytical approach are valid (this is performed a posteriori by examining the obtained numerical results).
To this end, we consider three different combinations of physical mass ratios $\pmr$ and damping ratios $\D$ as specified in the legend of \fref{verify}.
Recall from \ssref{prediction} that the physical mass ratio is defined as $\pmr={\ma}/{m_\mrm s}$, where the mass of the host structure is here defined as $m_\mrm s=\rho_\mrm s hb\ell$.
We vary the clearance in a wide range. 
For each parameter set, we compute the steady-state frequency response as described above and extract the corresponding resonant response.
From the resonant response, we extract the mean amplitude, the average number of impacts per period, the mean contact phase angle and the mean modal coefficient of restitution (as defined in \eref{corsim}).
The results are depicted in \fref{verify}.
\begin{figure}[b!]
	\centering
	\includegraphics[width=0.91\textwidth]{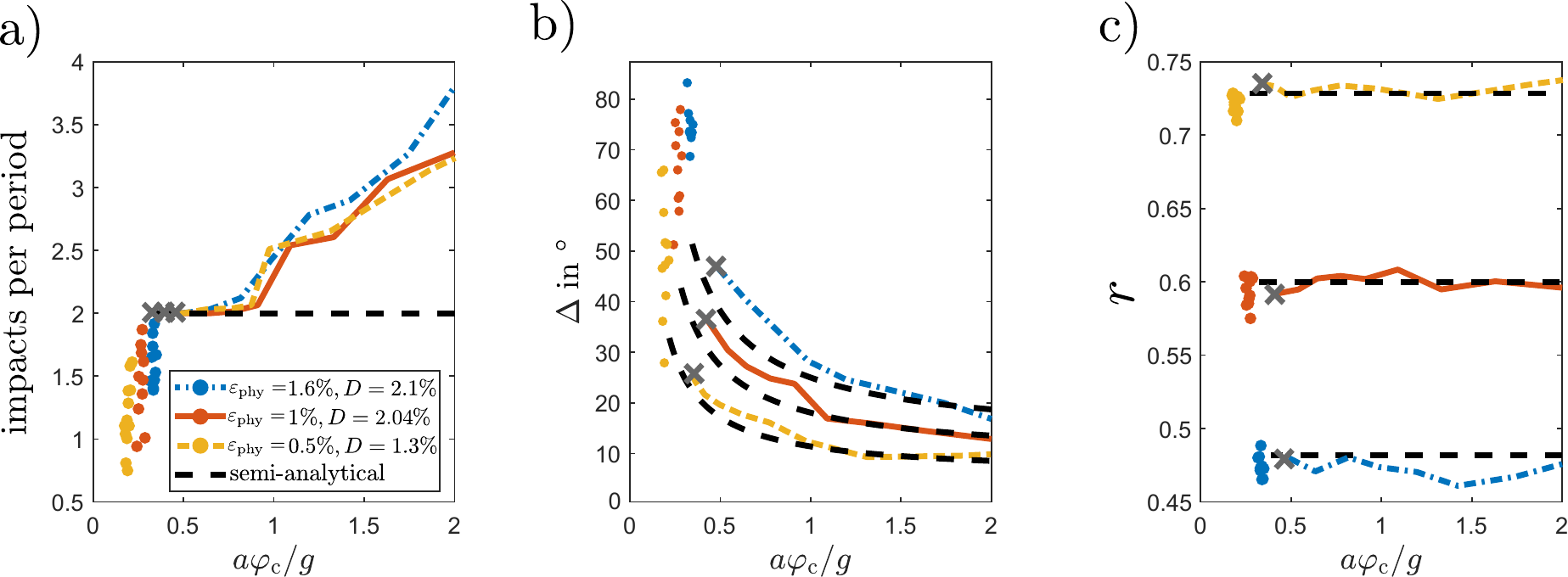}
	\caption{Numerical assessment of the basic assumptions of the semi-analytical approach: (a) Number of impacts per period, (b) contact phase angle, (c) coefficient of restitution; dots indicate Strongly Modulated Responses, lines indicate (almost) periodic responses, gray $\times$ indicates transition between Strongly Modulated Response and periodic response; results are obtained for varying clearance as described in the text.}
	\label{fig:verify}
\end{figure}
\\
As can be seen from \fref{verify}a, the assumption of on average two impacts per period is valid near the transition (marked with gray $\times$) between Strongly Modulated Response and (nearly) periodic response.
As expected, the number of impacts per period is lower than two in the Strongly Modulated Response regime, and increases beyond two for sub-optimal clearances in the almost periodic regime. 
As seen from \fref{verify}b, the amplitude-phase relation according to \eref{apsir} (black dashed lines) is in good agreement with the numerical reference near the optimum design point.
This supports the assumption that the contact instants and pre-impact relative velocity are mainly imposed by the resonant mode (and are independent of off-resonant modes).
As seen from \fref{verify}c, the coefficient of restitution varies only weakly with the clearance (and thus amplitude), and the values predicted for the optimum design point are in good agreement with the numerical reference. These results validate the proposed semi-analytical approach.\\

\subsection{Validation of the frequency response and efficacy curve} \label{sec:frfeff}
We now set the mass ratio equal to $\pmr=1\%$ and the modal damping ratio to $\D=2.04\%$, and vary the clearance in a wide range.
The resulting resonant response level and the frequency responses for selected, representative clearances are depicted in \fref{frfeff}.
\begin{figure}[b!]
	\centering
	\includegraphics[width=0.99\textwidth]{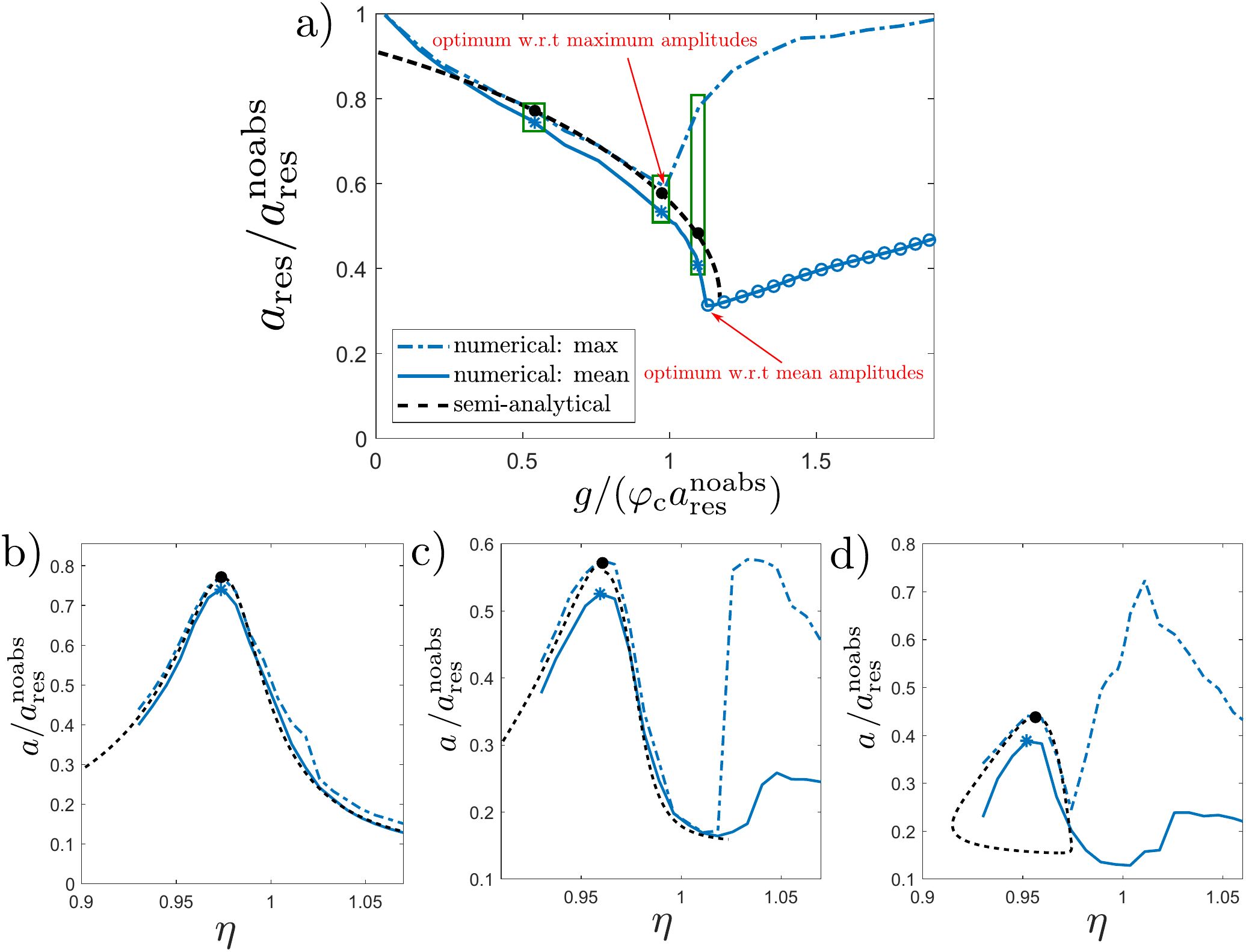}
	\caption{Validation of the semi-analytical approach against numerical reference: (a) Normalized resonant amplitude vs. normalized clearance, and (b),(c),(d) frequency response for $\gap=0.53\phic \aresnoabs$, $\gap=0.99\phic \aresnoabs$ and $\gap=1.08\phic \aresnoabs$, respectively, for $\pmr=1\%, \D=2.04\%$; circles in (a) indicate that the resonant response is strongly modulated.}
	\label{fig:frfeff}
\end{figure}
\\
The predicted amplitudes are in very good agreement with the numerically obtained mean response level in the regime of the almost periodic response.
Also, the optimum clearance and the associated resonant amplitude are in very good agreement.
The results deviate for smaller clearances, namely $\gap/(\phic\aresnoabs)\lessapprox 0.25$, as the number of impacts is then much larger than two, in full accordance with \fref{verify}a. This deviation is to be expected given the assumption of two impacts per period in our previous analysis.
\\
One should remark that throughout this work, we focused on the design objective of minimizing the \emph{mean} amplitudes.
As can be seen in \fref{frfeff}a, if the design objective is to minimize the \emph{maximum} amplitudes, one should select a smaller clearance and must compromise with a less pronounced vibration reduction effect.
This is to be expected since the maximum amplitudes in the Strongly Modulated Response regime can be much larger than the mean amplitudes.
As the proposed approach relies on the assumption of a strictly periodic response, we unfortunately do not have any information on the mean or maximum amplitudes reached in the Strongly Modulated Response regime.

\subsection{Validation of the optimum design point as function of mass ratio and damping ratio} \label{sec:depend}
Lastly, we focus on the optimum design point (with respect to the mean amplitude) and vary the mass ratio and the damping ratio in a wide range.
The results are depicted in \fref{depend} in terms of the optimum normalized clearance and the associated normalized resonant response level.
\begin{figure}[b!]
	\centering
	\includegraphics[width=0.76\textwidth]{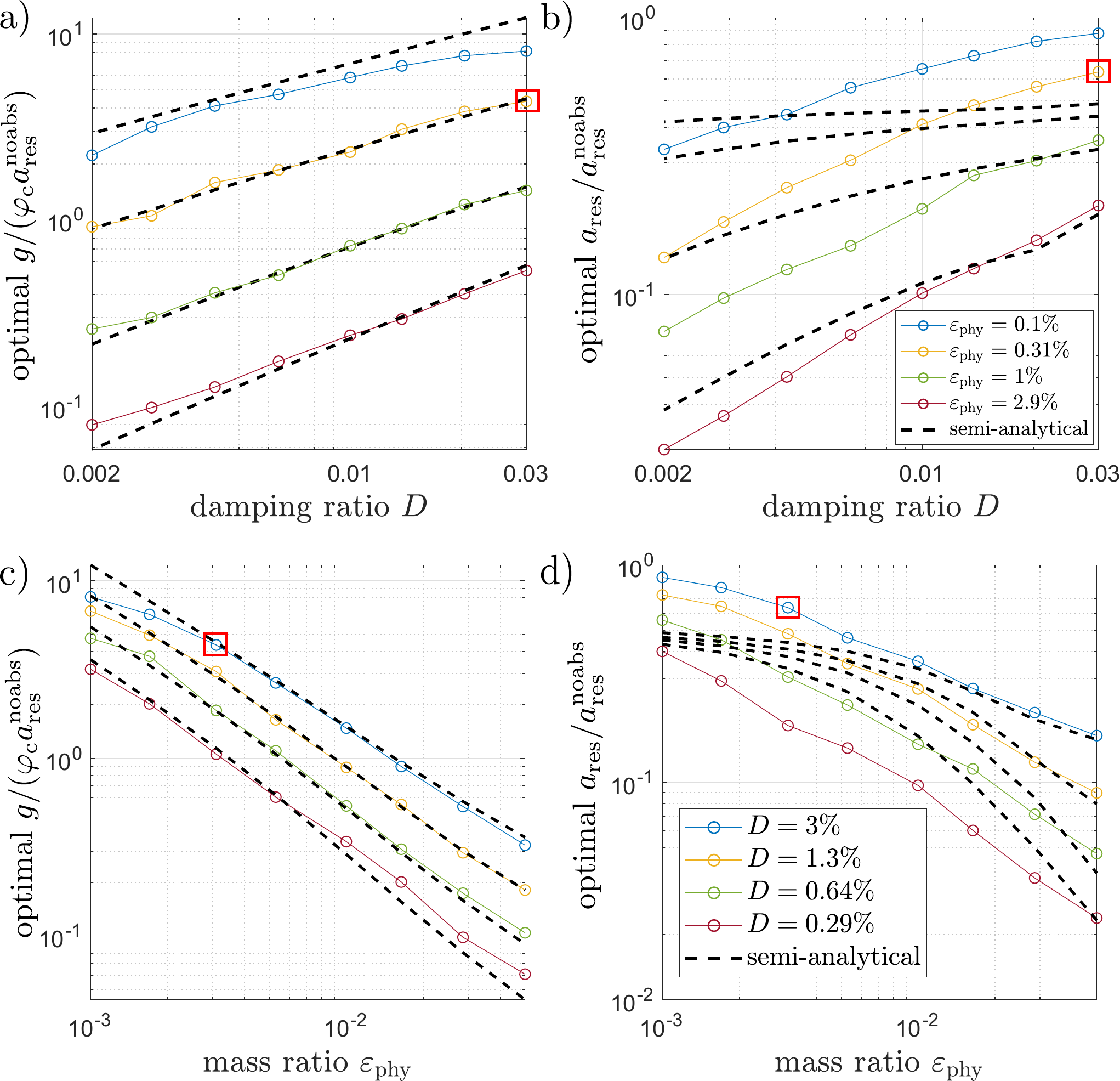}
	\caption{Optimal normalized clearance (left column) and associated normalized resonant response (right column) with respect to the mean amplitude as function of (a)-(b) the damping ratio, and (c)-(d) the mass ratio. The red squares indicate a representative parameter combination further analyzed in \fref{discrep}.}
	\label{fig:depend}
\end{figure}
%
The optimum clearance is predicted with very good accuracy over a wide range of the parameters.
Slight discrepancies can be observed for very high damping, and for very light damping in combination with high mass ratio.
The optimum efficacy (minimum resonant amplitude normalized by the corresponding response without absorber), as shown in the right column of \fref{depend}, is less well predicted.
The reason for this is the vertical tangent at the optimum design point (\fref{frfeff}a), leading to an extremely high sensitivity with respect to small errors.
The semi-analytical approach developed in this work is therefore useful to design the absorber for a given host structure and excitation scenario.
In order to assess the actual performance of the selected design, however, we recommend to verify/improve the semi-analytical prediction by direct numerical simulation.
\\
To further analyze the discrepancy between the semi-analytical prediction and the numerical reference, we consider a representative parameter combination, that is $\pmr=0.31\%$ and $D=3\%$, as marked by the red squares in \fref{depend}.
The resonant amplitude vs. normalized clearance, along with a few frequency responses, are depicted in \fref{discrep}.
\begin{figure}[b!]
	\centering
	\includegraphics[width=0.99\textwidth]{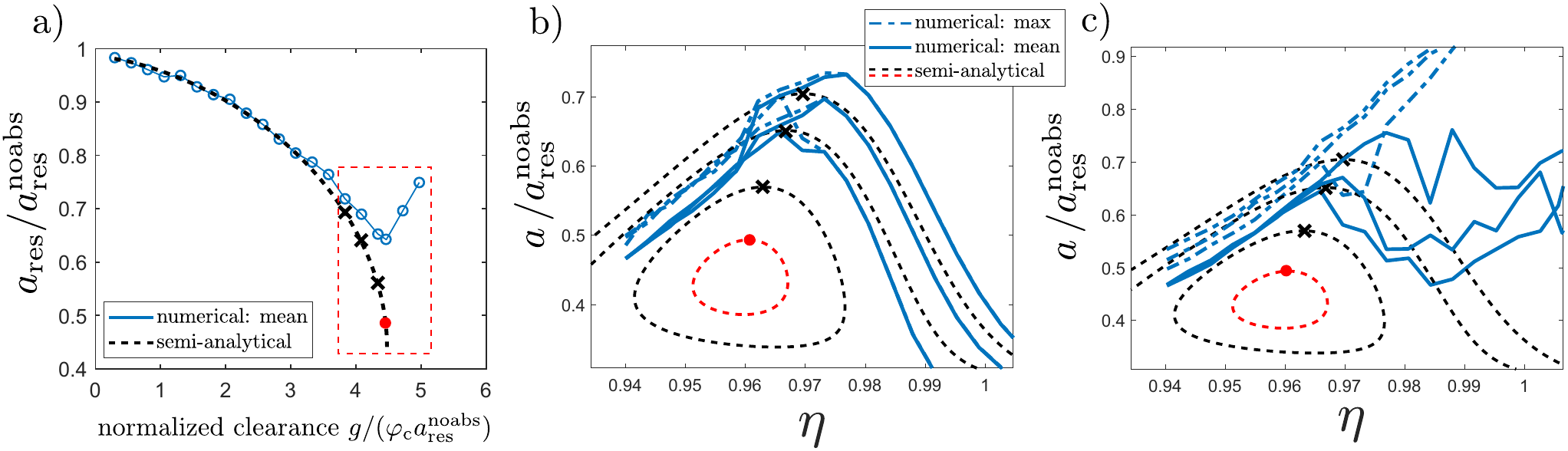}
	\caption{Towards the explanation of the discrepancy between semi-analytical prediction and numerical reference: (a) Efficacy curve, (b)-(c) frequency responses near the optimum design point for normalized clearances $\gap/(\phic\aresnoabs) \in \lbrace 3.8, 4.1, 4.3 \rbrace$ and $\gap/(\phic\aresnoabs) \in \lbrace 4.45, 4.7, 5 \rbrace$, respectively. For orientation, the semi-analytically predicted frequency responses (black/red dashed lines) for normalized clearances $\gap/(\phic\aresnoabs) \in \lbrace 3.8, 4.1, 4.3, 4.45 \rbrace$ are shown both in (b) and (c). $\pmr=0.31\%$, $D=3\%$.}
	\label{fig:discrep}
\end{figure}
%
In the frequency response diagrams, it can be seen that the resonant response (associated with the maximum mean amplitude) at and near the optimum design point, is in fact strongly modulated.
This violates the well-established hypothesis that the (almost) periodic response remains the critical one until it ceases to exist \cite{Li.2016b}.
Apparently, this hypothesis is not valid in an exact way, but only in reasonable approximation.

%

\section{Summary and conclusions} \label{sec:concl}
In this work, we developed a predictive approach to design impact absorbers for mitigating resonant vibrations.
In line with previous works, we focused on the periodic response regime with two symmetric impacts per period.
For the case of a well-separated resonance, we exploited the fact that the response is dominated by the fundamental harmonic of the resonant mode.
Under these assumptions, the absorber (approximately) oscillates in the form of a sawtooth function and our approach recovers the Slow Invariant Manifold well-known for the case of single-degree-of-freedom host structures.
In contrast to previous analytical studies, our approach is applicable to flexible host structures.
The truncation to the resonant mode gives rise to the notion of a modal coefficient of restitution.
Moreover, we have shown how this coefficient can be predicted numerically, by properly resolving the contact mechanics and the elasto-dynamics of a single, representative impact event.
This short-time high-fidelity simulation ensures a quantitatively accurate prediction of the optimum design, but renders the approach semi-analytical.
From this simulation, we also obtained parameters defining the shape of the contact force pulse, needed to set up the averaged equations governing the frequency response.
The analytical treatment permitted the reduction of the problem to just four dimensionless parameters: The modal coefficient of restitution, the damping ratio of the resonant mode, the static modal deflection at the absorber location divided by the clearance, and the modal mass ratio multiplied by a shape factor of the contact force pulse.
Furthermore, we derived a closed-form expression for the resonant amplitude as function of the clearance.
\\
The comparison of the semi-analytical approach against direct numerical simulation showed that the optimum design is predicted with very high accuracy.
However, only moderate accuracy with regard to the resulting vibration level was achieved, so that we propose to numerically recalculate the performance once a suitable design is selected.
The numerical results also indicated that the hypothesis that the (almost) periodic response remains critical until it ceases to exist, only holds approximately.
Finally, it should be remarked that the design objective in this work was to minimize the mean amplitudes.
As highlighted in the numerical results, a smaller clearance should be selected if the design objective is to minimize the maximum amplitudes.
To predictively design towards this objective, one would have to determine the maximum amplitude reached in the Strongly Modulated Response regime, a task, which lies outside the scope of the present work.

%

\section*{Acknowledgments}
The authors are grateful to MTU Aero Engines AG for providing the financial support for this work and for giving permission to publish it. Alexander F. Vakakis is grateful to the Alexander Humboldt Foundation for funding his visit at the University of Stuttgart through a Research Award.

\appendix
\setcounter{figure}{0}
\setcounter{table}{0}

\section{Relation between the amplitude $\amp$ and the contact phase angle $\psiA$ via the method of Non-Smooth Temporal Transformation (NSTT)}
Following closely the analysis in \cite{Pilipchuk.2015}, the method of Non-Smooth Temporal Transformation (NSTT) can be employed to alternatively obtain relation \ref{eq:apsir} between the unknown amplitude $\amp$ and contact phase angle $\psiA$. The approach outlined in \cite{Pilipchuk.2015} is applied to an impact absorber responding to the harmonic motion of the host structure at the contact location $\mm w\tra \mm q_\mrm{s}(t) = \phic\amp\sin(\Omega t + \psiA)$ as described in \sref{assumptions}.\\
The general idea behind the method of NSTT is the projection of the system's dynamics onto a set of non-smooth functions $\tau(t), e(t)$ generated by an impact absorber moving between rigid barriers. The function $\tau(t)$ can be viewed as the new temporal variable given by the triangle wave (or sawtooth) function $\tau(t) = 2/\pi\arcsin\left(\sin\left(\pi t/2\right)\right)$, and $e(t)$ and $\dot{e}(t)$ are its 1st and 2nd generalized derivatives with respect to time $t$. Derivatives with respect to the time variable $\tau$ are labeled with $(.)'$. The idea behind this transformation is that any periodic process $z(t)$ with normalized period $T=4$ can be expressed in terms of the new basis $\lbrace\tau,e\rbrace$ in form of the \emph{hyperbolic complex combination} $z(t)=X(\tau)+Y(\tau)e$ with $e^2=1$. In the specific example here, the transformation is applied to the relative motion
\begin{align} \label{eq:traforel}
	\q_\mrm{rel}(t)=\q_\mrm{a}(t)-\mm w\tra \mm q_\mrm{s}(t)=X(\tau(\phi))+Y(\tau(\phi))e(\phi)\fk\\
	\text{with phase } \phi(t) = \bar{\omega} t; \;\bar{\omega} = \frac{2\Omega}{\pi} = \text{const.}; \,\dot{\bar{\omega}}=0.\nonumber
\end{align}
The equation of motion of the system $\ma \ddot{q}_\mrm{a} = 0$ together with the constraints $|\q_\mrm{rel}|\leq \gap$ are replaced by the effective model without constraints,
\begin{align}
	\ma \ddot{\q}_\mrm{a} = \ma (\ddot{\q}_\mrm{rel}+\mm w\tra\ddot{\mm q}_\mrm{s}(t)) &= pe'\nonumber\\
	\ma \ddot{\q}_\mrm{rel} &= pe' + \ma \phic\amp\Omega^2\sin(\Omega t + \psiA)\fk
\end{align}
where the reaction of the constraints is replaced by a periodic series of Dirac impulses $pe'$ of unknown amplitude $p$. Considering the hyperbolic complex representation of the cavity motion,
\begin{equation}
	\sin(\Omega t + \psiA) = R_\mrm f + I_\mrm f e = \cd\sin\left(\frac{\pi}{2}\tau\right) + \sd\cos\left(\frac{\pi}{2}\tau\right)e\fk
\end{equation}
together with the 2nd derivative of \eref{traforel} with respect to time $t$,
\begin{equation}
	\ddot{\q}_\mrm{rel} = X''\bar{\omega}^2 + Y'\cancel{\dot{\bar{\omega}}} + \left(Y''\bar{\omega}^2 + X'\cancel{\dot{\bar{\omega}}}\right)e + X'e'\bar{\omega}^2\; ,
\end{equation}
we get the transformed equation:
\begin{equation}
	\ma \bar{\omega}^2 \left(X'' + Y''e + X'e'\right) = pe' + \ma \phic\amp \Omega^2\left(\cd\sin\left(\frac{\pi}{2}\tau\right) + \sd\cos\left(\frac{\pi}{2}\tau\right)e\right)\,.
\end{equation}
Setting equal to zero separately the coefficients of $\lbrace1,e,e'\rbrace$ yields
\begin{align}
	&1:\quad \bar{\omega}^2X'' = \phic\amp\Omega^2\cd\sin\left(\frac{\pi}{2}\tau\right)\label{eq:ODE1}\fk\\
	&e:\quad \bar{\omega}^2Y'' = \phic\amp\Omega^2\sd\cos\left(\frac{\pi}{2}\tau\right)\label{eq:ODE2}\fk\\
	&\text{and boundary condition (BC)}\nonumber\\
	&e':\quad \text{at } \tau = \pm 1: \ma\bar{\omega}^2 X' = p \,.\label{eq:BVPBC}
\end{align}
We note that in contrast to the original model, equations \eref{ODE1} and \eref{ODE2} are smooth (\ie they do not contain any terms with singularities), and in the form of boundary value problems (BVPs) with boundary conditions given by \eref{BVPBC}. What is even more convenient is that analytical approximations of the solutions of these BVPs can be obtained in the form of regular perturbation series in the variable $\tau$. Indeed, the general solutions of the ordinary differential equations \ref{eq:ODE1} and \ref{eq:ODE2} are
\begin{align}
	X(\tau) = F\tau + G - \frac{\Omega^2}{\bar{\omega}^2}\phic\amp\cd\sin\left(\frac{\pi}{2}\tau\right)\frac{4}{\pi^2}\fk\\
	Y(\tau) = H\tau + J - \frac{\Omega^2}{\bar{\omega}^2}\phic\amp\sd\cos\left(\frac{\pi}{2}\tau\right)\frac{4}{\pi^2}\fk
\end{align}
with boundary conditions
\begin{subnumcases}{\text{at } \tau = \pm 1}
	X = \pm \gap\label{eq:BC1}\\
	Y = 0\label{eq:BC2}\\
	Y'\mp X' = -\cor\left(Y' \pm X'\right)\label{eq:BC3}\\
	\ma X\bar{\omega}^2 = p,\label{eq:BC4}
\end{subnumcases}
where $\cor$ is the coefficient of restitution.
From \eref{BC1} and \eref{BC2} we get $F=\frac{\Omega^2}{\bar{\omega}^2}\phic\amp\cd\frac{4}{\pi^2}+\gap,\, G=0$ and $H=J=0$. From \eref{BC3} with $\tau=1$ (or $\tau=-1$) and together with $\frac{\Omega}{\bar{\omega}}=\frac{\pi}{2}$ we find
\begin{align}
	\frac{\Omega^2}{\bar{\omega}^2}\phic\amp \sd\frac{2}{\pi}-\frac{\Omega^2}{\bar{\omega}^2}\phic\amp \cd\frac{4}{\pi^2}-\gap&= -\cor\left(\frac{\Omega^2}{\bar{\omega}^2}\phic\amp \sd \frac{2}{\pi}+\frac{\Omega^2}{\bar{\omega}^2}\phic\amp \cd\frac{4}{\pi^2}+\gap\right)\nonumber\\
	\frac{\pi}{2}\phic\amp\left(\sd(1+\cor)-\frac{2}{\pi}(1-\cor)\cd\right)&=\gap(1-\cor)\fk
\end{align}
and finally
\begin{equation} \label{eq:apsir2}
	\frac{\gap}{\phic\amp}=\frac{\pi}{2}\frac{1+\cor}{1-\cor}\sd-\cd\, ,
\end{equation}
which is \eref{apsir} in the main text.

\section*{References}
\bibliography{literature_mk}

\begin{thebibliography}{10}
\expandafter\ifx\csname url\endcsname\relax
  \def\url#1{\texttt{#1}}\fi
\expandafter\ifx\csname urlprefix\endcsname\relax\def\urlprefix{URL }\fi
\expandafter\ifx\csname href\endcsname\relax
  \def\href#1#2{#2} \def\path#1{#1}\fi

\bibitem{Hartung2016}
A.~Hartung, U.~Retze, H.-P. Hackenberg, {Impulse Mistuning of Blades and
  Vanes}, Proceedings of ASME Turbo Expo 2016: Turbomachinery Technical
  Conference and Exposition (2016) 1--10\href
  {http://dx.doi.org/10.1115/GT2016-56433} {\path{doi:10.1115/GT2016-56433}}.

\bibitem{vaka2008b}
A.~F. Vakakis, O.~V. Gendelman, G.~Kerschen, L.~A. Bergman, D.~M. McFarland,
  Y.~S. Lee, Nonlinear Targeted Energy Transfer in Mechanical and Structural
  Systems, {Springer Berlin Heidelberg}, 2008.

\bibitem{Lamarque2011}
C.-H. Lamarque, O.~V. Gendelman, A.~{Ture Savadkoohi}, E.~Etcheverria, Targeted
  energy transfer in mechanical systems by means of non-smooth nonlinear energy
  sink, Acta Mechanica 221~(1) (2011) 175.
\newblock \href {http://dx.doi.org/10.1007/s00707-011-0492-0}
  {\path{doi:10.1007/s00707-011-0492-0}}.

\bibitem{Gendelman.2012}
O.~V. Gendelman, Analytic treatment of a system with a vibro-impact nonlinear
  energy sink, Journal of Sound and Vibration 331~(21) (2012) 4599--4608.
\newblock \href {http://dx.doi.org/10.1016/j.jsv.2012.05.021}
  {\path{doi:10.1016/j.jsv.2012.05.021}}.

\bibitem{Lieber.1945}
P.~Lieber, D.~Jensen, et~al., An acceleration damper: development, design and
  some applications, Trans. ASME 67~(10) (1945) 523--530.

\bibitem{AlShudeifat.2013}
M.~A. Al-Shudeifat, N.~Wierschem, D.~D. Quinn, A.~F. Vakakis, L.~A. Bergman,
  B.~F. {Spencer Jr.}, Numerical and experimental investigation of a highly
  effective single-sided vibro-impact non-linear energy sink for shock
  mitigation, International Journal of Non-Linear Mechanics 52 (2013) 96--109.
\newblock \href {http://dx.doi.org/10.1016/j.ijnonlinmec.2013.02.004}
  {\path{doi:10.1016/j.ijnonlinmec.2013.02.004}}.

\bibitem{Bapat.1985}
C.~N. Bapat, S.~Sankar, Single unit impact damper in free and forced vibration,
  Journal of Sound and Vibration 99~(1) (1985) 85--94.
\newblock \href {http://dx.doi.org/10.1016/0022-460X(85)90446-8}
  {\path{doi:10.1016/0022-460X(85)90446-8}}.

\bibitem{Gendelman.2015}
O.~V. Gendelman, A.~Alloni, Dynamics of forced system with vibro-impact energy
  sink, Journal of Sound and Vibration 358 (2015) 301--314.
\newblock \href {http://dx.doi.org/10.1016/j.jsv.2015.08.020}
  {\path{doi:10.1016/j.jsv.2015.08.020}}.

\bibitem{Theurich.2019}
T.~Theurich, J.~Gross, M.~Krack, Effects of modal energy scattering and
  friction on the resonance mitigation with an impact absorber, Journal of
  Sound and Vibration 442 (2019) 71--89.
\newblock \href {http://dx.doi.org/10.1016/j.jsv.2018.10.055}
  {\path{doi:10.1016/j.jsv.2018.10.055}}.

\bibitem{Chatterjee.1996}
S.~Chatterjee, A.~K. Mallik, A.~Ghosh, {Impact dampers for controlling
  self-excited oscillation}, Journal of Sound and Vibration 193~(5) (1996)
  1003--1014.
\newblock \href {http://dx.doi.org/10.1006/jsvi.1996.0327}
  {\path{doi:10.1006/jsvi.1996.0327}}.

\bibitem{Asfar.2005}
K.~R. Asfar, S.~N. Akour, {Optimization analysis of impact viscous damper for
  controlling self-excited vibrations}, JVC/Journal of Vibration and Control
  11~(1) (2005) 103--120.
\newblock \href {http://dx.doi.org/10.1177/1077546305048325}
  {\path{doi:10.1177/1077546305048325}}.

\bibitem{mueller.2018}
P.~M{\"u}ller, A.~Hartung, H.-P. Hackenberg, A study of dynamic phenomena
  caused by impulse mistuning in case of self-excitation, in: Turbo Expo: Power
  for Land, Sea, and Air, Vol. 51159, American Society of Mechanical Engineers,
  2018, p. V07CT35A004.

\bibitem{Li.2017}
T.~Li, E.~Gourc, S.~Seguy, A.~Berlioz, Dynamics of two vibro-impact nonlinear
  energy sinks in parallel under periodic and transient excitations,
  International Journal of Non-Linear Mechanics 90 (2017) 100--110.
\newblock \href {http://dx.doi.org/10.1016/j.ijnonlinmec.2017.01.010}
  {\path{doi:10.1016/j.ijnonlinmec.2017.01.010}}.

\bibitem{Boroson.2017}
E.~Boroson, S.~Missoum, P.-O. Mattei, C.~Vergez, Optimization under uncertainty
  of parallel nonlinear energy sinks, Journal of Sound and Vibration 394 (2017)
  451--464.
\newblock \href {http://dx.doi.org/10.1016/j.jsv.2016.12.043}
  {\path{doi:10.1016/j.jsv.2016.12.043}}.

\bibitem{Vaurigaud.2011}
B.~Vaurigaud, A.~{Ture Savadkoohi}, C.-H. Lamarque, Targeted energy transfer
  with parallel nonlinear energy sinks. part i: Design theory and numerical
  results, Nonlinear Dynamics 66~(4) (2011) 763--780.
\newblock \href {http://dx.doi.org/10.1007/s11071-011-9949-x}
  {\path{doi:10.1007/s11071-011-9949-x}}.

\bibitem{Fang.2020}
B.~Fang, T.~Theurich, M.~Krack, L.~A. Bergman, A.~F. Vakakis, Vibration
  suppression and modal energy transfers in a linear beam with attached
  vibro-impact nonlinear energy sinks, Communications in Nonlinear Science and
  Numerical Simulation 91 (2020) 105415.
\newblock \href {http://dx.doi.org/10.1016/j.cnsns.2020.105415}
  {\path{doi:10.1016/j.cnsns.2020.105415}}.

\bibitem{Qiu.2019}
D.~Qiu, S.~Seguy, M.~Paredes, Design criteria for optimally tuned vibro-impact
  nonlinear energy sink, Journal of Sound and Vibration 442 (2019) 497--513.
\newblock \href {http://dx.doi.org/10.1016/j.jsv.2018.11.021}
  {\path{doi:10.1016/j.jsv.2018.11.021}}.

\bibitem{Li.2016b}
T.~Li, S.~Seguy, A.~Berlioz, Optimization mechanism of targeted energy transfer
  with vibro-impact energy sink under periodic and transient excitation,
  Nonlinear Dynamics (2016) 1--19\href
  {http://dx.doi.org/10.1007/s11071-016-3200-8}
  {\path{doi:10.1007/s11071-016-3200-8}}.

\bibitem{Gourc.2015}
E.~Gourc, G.~Michon, S.~Seguy, A.~Berlioz, Targeted energy transfer under
  harmonic forcing with a vibro-impact nonlinear energy sink: Analytical and
  experimental developments, Journal of Vibration and Acoustics 137~(3) (2015)
  031008--031008--7.
\newblock \href {http://dx.doi.org/10.1115/1.4029285}
  {\path{doi:10.1115/1.4029285}}.

\bibitem{Li.2016}
T.~Li, S.~Seguy, A.~Berlioz, On the dynamics around targeted energy transfer
  for vibro-impact nonlinear energy sink, Nonlinear Dynamics (2016)\href
  {http://dx.doi.org/10.1007/s11071-016-3127-0}
  {\path{doi:10.1007/s11071-016-3127-0}}.

\bibitem{Li.2017b}
T.~Li, D.~Qiu, S.~Seguy, A.~Berlioz, Activation characteristic of a
  vibro-impact energy sink and its application to chatter control in turning,
  Journal of Sound and Vibration 405 (2017) 1--18.
\newblock \href {http://dx.doi.org/10.1016/j.jsv.2017.05.033}
  {\path{doi:10.1016/j.jsv.2017.05.033}}.

\bibitem{Luo.2014}
J.~Luo, N.~E. Wierschem, S.~A. Hubbard, L.~A. Fahnestock, D.~{Dane Quinn},
  D.~{Michael McFarland}, B.~F. Spencer, A.~F. Vakakis, L.~A. Bergman,
  Large-scale experimental evaluation and numerical simulation of a system of
  nonlinear energy sinks for seismic mitigation, Engineering Structures 77
  (2014) 34--48.
\newblock \href {http://dx.doi.org/10.1016/j.engstruct.2014.07.020}
  {\path{doi:10.1016/j.engstruct.2014.07.020}}.

\bibitem{AlShudeifat.2015}
M.~A. Al-Shudeifat, A.~F. Vakakis, L.~A. Bergman, Shock mitigation by means of
  low- to high-frequency nonlinear targeted energy transfers in a large-scale
  structure, Journal of Computational and Nonlinear Dynamics 11~(2) (2015)
  1--11.

\bibitem{Pilipchuk.2015}
V.~N. Pilipchuk, Closed-form solutions for oscillators with inelastic impacts,
  Journal of Sound and Vibration 359 (2015) 154--167.
\newblock \href {http://dx.doi.org/10.1016/j.jsv.2015.08.023}
  {\path{doi:10.1016/j.jsv.2015.08.023}}.

\bibitem{hert1881}
H.~Hertz, Ueber die ber{\"u}hrung fester elastischer k{\"o}rper, Journal
  f{\"u}r die reine und angewandte Mathematik 92 (1881) 156--171.

\bibitem{john1989}
K.~L. Johnson, Contact Mechanics, {Cambridge University Press}, Cambridge,
  1989.

\bibitem{Seifried.2010}
R.~Seifried, W.~Schiehlen, P.~Eberhard, The role of the coefficient of
  restitution on impact problems in multi-body dynamics, Proceedings of the
  IMechE 224~(3) (2010) 279--306.
\newblock \href {http://dx.doi.org/10.1243/14644193JMBD239}
  {\path{doi:10.1243/14644193JMBD239}}.

\bibitem{Hunter.1957}
S.~C. Hunter, Energy absorbed by elastic waves during impact, Journal of the
  Mechanics and Physics of Solids 5~(3) (1957) 162--171.
\newblock \href {http://dx.doi.org/10.1016/0022-5096(57)90002-9}
  {\path{doi:10.1016/0022-5096(57)90002-9}}.

\bibitem{Reed.1985}
J.~Reed, Energy losses due to elastic wave propagation during an elastic
  impact, Journal of Physics D: Applied Physics 18~(12) (1985) 2329--2337.
\newblock \href {http://dx.doi.org/10.1088/0022-3727/18/12/004}
  {\path{doi:10.1088/0022-3727/18/12/004}}.

\end{thebibliography}

\end{document}